\documentclass[twocolumn,trackchanges]{aastex631}
\pdfoutput=1 
\usepackage[T1]{fontenc}
\usepackage[figure,figure*]{hypcap}
\usepackage{amsmath,amstext,bm}
\usepackage{graphicx,textcomp,fancyhdr,hyperref}

\DeclareRobustCommand{\okina}{%
\raisebox{\dimexpr\fontcharht\font`A-\height}{%
    \scalebox{0.8}{`}%
  }%
}

\shorttitle{KPF Solar Calibrator}
\shortauthors{Rubenzahl et al.}

\newcommand{\ms}{m~s$^{-1}$}
\newcommand{\cms}{cm~s$^{-1}$}

\newcommand{\Wmsq}[0]{W~m$^{-2}$}
\newcommand{\hjd}{HJD$_\text{TDB}$}
\newcommand{\bjd}{BJD$_\text{TDB}$}

\begin{document}

\title{Staring at the Sun with the Keck Planet Finder: An Autonomous Solar Calibrator\\for High Signal-to-Noise Sun-as-a-Star Spectra}

\newcommand{\caltech}{California Institute of Technology, Pasadena, CA 91125, USA}
\newcommand{\Caltechastro}{Department of Astronomy, California Institute of Technology, Pasadena, CA 91125, USA}
\newcommand{\JPL}{Jet Propulsion Laboratory, California Institute of Technology, 4800 Oak Grove Drive, Pasadena, CA 91109}
\newcommand{\WMKO}{W. M. Keck Observatory, 65-1120 Mamalahoa Hwy, Kamuela, HI, 96743}
\newcommand{\ipac}{NASA Exoplanet Science Institute/Caltech-IPAC, MC 314-6, 1200 E. California Blvd., Pasadena, CA 91125, USA}

\author[0000-0003-3856-3143]{Ryan A. Rubenzahl}
\altaffiliation{NSF Graduate Research Fellow}
\affil{\Caltechastro}

\author[0000-0003-1312-9391]{Samuel Halverson} 
\affil{\JPL}

\author[0000-0002-6092-8295]{Josh Walawender}
\affil{\WMKO} 

\author[0000-0002-7648-9119]{Grant M. Hill}
\affil{\WMKO} 
\author[0000-0001-8638-0320]{Andrew W. Howard}
\affil{\Caltechastro} 

\author[0000-0001-6516-6915]{Matthew Brown}
\affil{\WMKO} 

\author{Evan Ida}
\affil{Hawai\okina i Community College, 1175 Manono St, Hilo, HI 96720}
\affil{\WMKO} 

\author{Jerez Tehero}
\affil{\WMKO} 

\author[0000-0003-3504-5316]{Benjamin J. Fulton}
\affiliation{\ipac} 

\author{Steven R. Gibson}
\affiliation{\caltech} 

\author{Marc Kassis}
\affil{\WMKO} 

\author{Brett Smith}
\affil{\WMKO}  

\author{Truman Wold}
\affil{\WMKO} 

\author{Joel Payne}
\affil{\WMKO} 
\correspondingauthor{Ryan Rubenzahl}
\email{rrubenza@caltech.edu}

\begin{abstract}

Extreme precision radial velocity (EPRV) measurements contend with internal noise (instrumental systematics) and external noise (intrinsic stellar variability) on the road to 10~{\cms} ``exo-Earth'' sensitivity. Both of these noise sources are well-probed using ``Sun-as-a-star'' RVs and cross-instrument comparisons. We built the Solar Calibrator (SoCal), an autonomous system that feeds stable, disc-integrated sunlight to the recently commissioned Keck Planet Finder (KPF) at the W. M. Keck Observatory. With SoCal, KPF acquires signal-to-noise~$\sim$~1200, R~=~98,000 optical (445--870~nm) spectra of the Sun in 5~sec exposures at unprecedented cadence for an EPRV facility using KPF's fast readout mode ($<16$~sec between exposures). Daily autonomous operation is achieved by defining an operations loop using state machine logic. Data affected by clouds are automatically flagged using a reliable quality control metric derived from simultaneous irradiance measurements. Comparing solar data across the growing global network of EPRV spectrographs with solar feeds will allow EPRV teams to disentangle internal and external noise sources and benchmark spectrograph performance. To facilitate this, all SoCal data products are immediately available to the public on the Keck Observatory Archive. We compared SoCal RVs to contemporaneous RVs from NEID, the only other immediately public EPRV solar dataset. We find agreement at the 30--40~\cms\ level on timescales of several hours, which is comparable to the combined photon-limited precision. Data from SoCal were also used to assess a detector problem and wavelength calibration inaccuracies associated with KPF during early operations. Long-term SoCal operations will collect upwards of 1,000 solar spectra per six-hour day using KPF's fast readout mode, enabling stellar activity studies at high signal-to-noise on our nearest solar-type star.

\end{abstract}

\section{Introduction}\label{sec:intro}

Since the first detection of an exoplanet with radial velocities (RVs), 51~Pegasi~b \citep[50~{\ms} semiamplitude;][]{MayorQueloz1995}, RV instruments can now detect RV signals as small as 50~{\cms} \citep[e.g.][]{Zhao2023_55cnce}. This leap of two orders of magnitude in sensitivity has been enabled by cycles of instrumentation development, rigorous testing, and a systematic understanding of the myriad instrumental systematics in modern and next-generation Extreme Precision Radial Velocity (EPRV) spectrographs \citep{Halverson2016}. There are now a number of such instruments with sub-{\ms} capability, including the Keck Planet Finder \citep[KPF;][]{Gibson2016,Gibson2018,Gibson2020}, the High-Accuracy Radial-velocity Planet Searcher \citep[HARPS;][]{HARPS} and its northern twin \citep[HARPS-N;][]{HARPSN}, the Echelle SPectrograph for Rocky Exoplanets and Stable Spectroscopic Observations \citep[ESPRESSO;][]{ESPRESSO, Pepe2021}, the EXtreme PREcision Spectrometer \citep[EXPRES;][]{EXPRES}, the NN-explore Exoplanet Investigations with Doppler spectroscopy instrument \citep[NEID;][]{NEID}, and the M dwarf Advanced Radial velocity Observer Of Neighboring eXoplanets \citep[MAROON-X;][]{MAROONX}. The task remains to reduce this \textit{internal} noise floor to below 10~{\cms}; this is the level needed to measure the masses of Earth-like planets in 1~AU orbits around Sun-like stars (9~{\cms} RV semiamplitude). In fact, the EPRV measurement technique remains the only viable method to make such a measurement \citep{eprv_working_group_report}, and is perhaps the most promising method for discovering exo-Earths for follow-up characterization by the future Habitable Worlds Observatory \citep{astro2020}.

To complicate the precision goal of the EPRV community, surface phenomena on stars can induce apparent RV variability of up to many m~s$^{-1}$ \citep{HaywoodThesisChapter2}. This \textit{external} noise is present across all timescales. Acoustic oscillations occur on timescales of minutes \citep{Kjeldsen2005, Arentoft2008, Dumusque2011a, Chaplin2019, Gupta2022}, while convective granulation \citep{DelMoro2004, Meunier2015, Dumusque2011a, Cegla2018} and supergranulation \citep{Rincon2018, Meunier2019} occur on hours--days timescales. Surface inhomogeneities (e.g. starspots, faculae, plage) which break the symmetry of the star's rotational velocity profile~\citep{Saar1997, Meunier2010-plages, Boisse2011, Dumusque2011b} as well as suppress the convective blueshift \citep[dominant source in the Sun;][]{Meunier2010, Meunier2013, Haywood2016, Milbourne2019} are modulated by the star's rotation period on weeks--months timescales. Long-term magnetic activity cycles can produce RV variations on decades timescales~\citep{Meunier2010-soho, Lovis2011, Luhn2022}. This so-called ``stellar activity'' can complicate the measurement of precise planetary properties \citep{Blunt2023}, mimic the signal of an exoplanet \citep{Lubin2021}, or otherwise prevent real planets from being detected, even with thousands of observations of a single star over decades \citep{Langellier2021, Luhn2023, Gupta2023}. While activity-induced RVs can be partially mitigated by intentional observing strategies \citep{Dumusque2011a}, algorithmic models \citep{Haywood2014, Rajpaul2015, Aigrain2023}, more robust RV extraction methods \citep{Dumusque2018}, or by detrending with ``activity indicators'' \citep{Queloz2009, Isaacson2010, Aigrain2012, Siegel2022}, it remains an active area of research to derive activity-\textit{invariant} RVs.

While stars as are observed as unresolved point sources, the surface of the Sun is under constant monitoring at multiple wavelengths in photometry, spectroscopy, polarimetry, and spectropolarimetry at high angular resolution from the ground \citep[e.g. The Daniel K. Inouye Solar Telescope][]{DKIST} and space \citep[e.g. NASA's Solar Dynamics Observatory][]{SDO}. There is no other star for which observed spectra and RVs can be studied in connection to directly observed active processes \citep[e.g.][]{Haywood2016, Thompson2020, Milbourne2021, Ervin2022} with full confidence that the observed RV variability is due to \textit{only} stellar and instrumental noise (i.e., all Solar System planets are known and their signals removed). This makes the Sun the ideal laboratory for studying how activity manifests in spectra, especially in solar-type stars, the primary target for discovering exo-Earths.

As such, solar feeds are becoming a crucial component of EPRV facilities. The first solar feed was the Low Cost Solar Telescope \citep[LCST;][]{Phillips2016} for HARPS-N at the Telescopio Nazionale Galileo (TNG), which has been observing the Sun-as-a-star since 2015. The HARPS Experiment for Light Integrated Over the Sun (HELIOS) was later added to HARPS in 2018, and the NEID Solar Feed \citep{Lin2022} and Lowell Observatory Solar Telescope (LOST; Llama in prep) both began operations in 2020. Also at TNG, the LOw Cost NIR Extended teleScope \citep[LOCNES;][]{LOCNES} was installed to feed GIANO-B \citep{GIANOB}. The Potsdam Echelle Polarimetric and Spectroscopic Instrument \citep[PEPSI;][]{PEPSI} also has a solar feed installed. Solar feeds are currently being installed for MAROON-X and Near Infra Red Planet Searcher \citep[NIRPS;][]{NIRPS}. The Paranal solar Espresso Telescope \citep[PoET;][]{poet_sensors} and A dual-Beam pOlarimetric Robotic Aperture for the Sun \citep[ABORAS;][]{ABORAS} are planned for ESPRESSO and HARPS-3 respectively.

Solar feeds can also be used to independently monitor the instrumental ``drift'' \citep{Lin2022}, diagnose instrumental problems, and perform commissioning tests without using (precious) telescope time at night.  These tests are often superior to tests with calibration sources because the stellar spectra are processed by the instrument's data reduction pipeline (DRP) in the same way as stellar spectra. Cross-comparisons between the various solar datasets are also uniquely advantageous. \citet{Zhao2023} compared one month of solar data between HARPS, HARPS-N, EXPRES, and NEID and found an astounding agreement of 15--30~{\cms} between instruments on intra-day timescales. Longer timescales showed a larger 50--60~{\cms} variability, but are more affected by unshared observing conditions (e.g. different differential extinction due to different airmasses and solar disk positions at each site at a given time). Importantly, common variability in such multi-instrument contemporaneous datasets can be uniquely attributed to astrophysical processes on the Sun, while variability seen in only one instrument can be diagnosed as intrinsic systematic noise. Of course, this requires multiple instruments to be on-Sun at the same time, which is complicated by the geographic location of each facility.

For all of these reasons, we designed and built the Solar Calibrator (SoCal) to feed disk-integrated sunlight to the Keck Planet Finder, a newly commissioned EPRV spectrograph at W. M. Keck Observatory. In Section~\ref{sec:design} we describe the design and hardware of SoCal. Section~\ref{sec:operations} details the daily operations procedure and autonomous control loop. We discuss the data reduction and quality control of the solar datastream in Section~\ref{sec:drp}. Lastly in Section~\ref{sec:results} we report on commissioning progress, present first results on the Sun, and validate KPF's performance as an EPRV facility.

\section{Instrument Design}\label{sec:design}

The Keck Planet Finder \citep[KPF;][]{Gibson2016,Gibson2018, Gibson2020} is a fiber-fed, ultra-stabilized EPRV system for the W. M. Keck Observatory (WMKO) that was recently commissioned in 2022. KPF is designed to achieve an instrumental measurement precision of $\sim$30~\cms{} or better. The KPF main spectrometer spans 445--870 nm in two separate channels with a median resolving power of 98,000, enabled by an image slicer assembly that slices the science fiber image into three separate channels. KPF is wavelength-calibrated by several sources including a commercial laser frequency comb from Menlo Systems, a broadband Fabry-P\'erot etalon, and hollow cathode lamps (ThAr and UNe). A simultaneous calibration fiber is used to track instantaneous instrumental drift, and a dedicated sky fiber is used to monitor background sky contamination. The core KPF spectrometer is designed around a novel all-Zerodur optical bench, which has a near-zero coefficient of thermal expansion to suppress instrumental systematics related to thermomechanical motions. KPF also includes a dedicated near-UV spectrometer to monitor the chromospheric Ca H\&K lines for stellar activity tracking. The combination of CCD pixels with deep wells and optical slicing of the science spectrum onto three traces spread out in cross-dispersion allows KPF to achieve per-spectrum signal-to-noise ratios (SNR) more than twice that of other EPRV facilities.

\begin{figure*}
\includegraphics[width=0.95\textwidth]{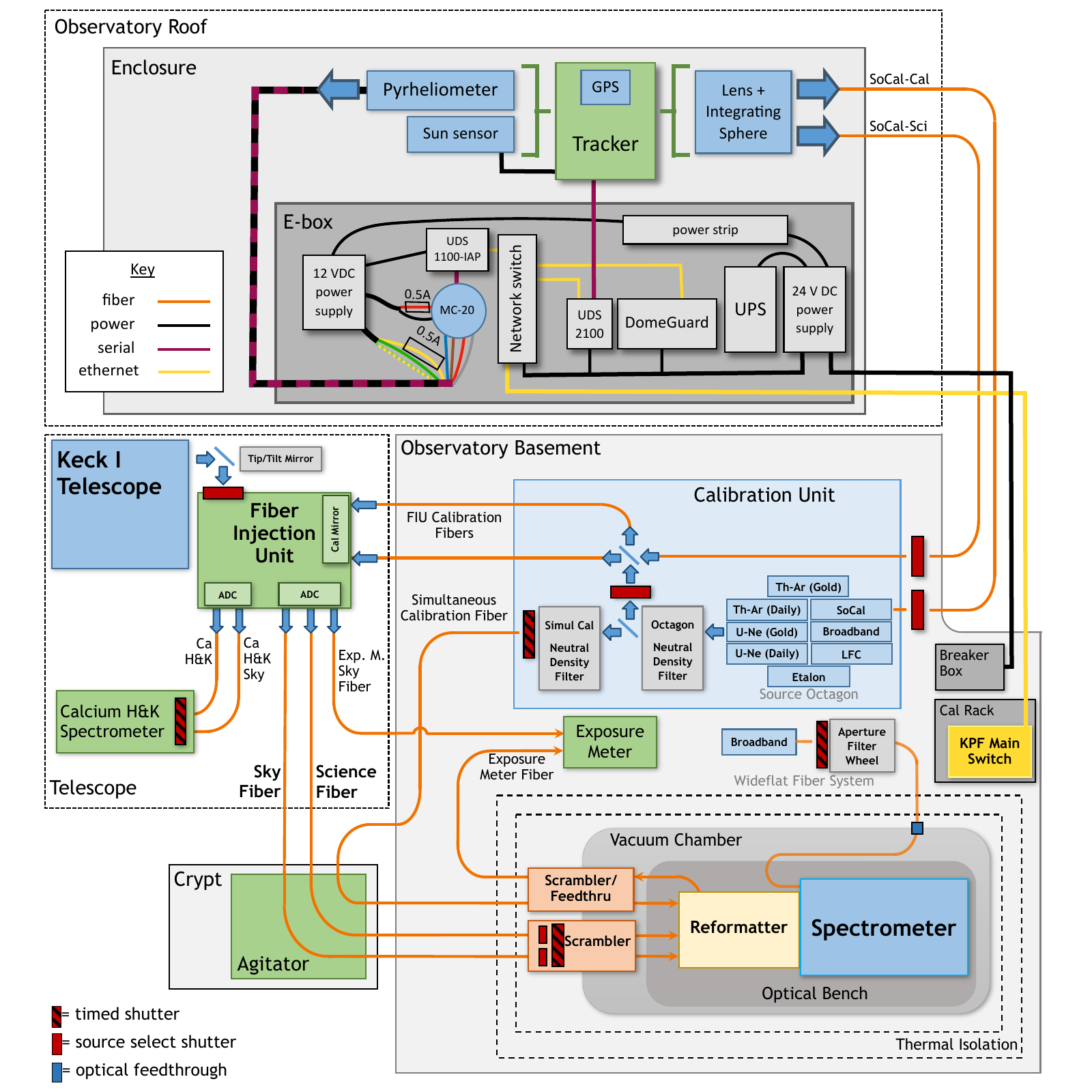}
\caption{Block diagram illustrating how SoCal interfaces with the rest of the KPF system.  SoCal is contained in the top dashed-line box labeled ``Observatory Roof.''  A pair of optical fibers carry sunlight to KPF's Calibration Unit in the observatory basement, where the ``SoCal-Cal'' fiber connects to the calibration source selector assembly and the ``SoCal-Sci'' fiber feeds dedicated calibration fibers that connect directly to the FIU. The latter path sends solar light through the same path as starlight from the Keck I telescope. In this mode, a calibration source (e.g., the etalon) can be used for simultaneous calibration. The pyrheliometer irradiance is directly recorded by a computer which polls every second. Also shown are the electronics in the SoCal electronics box (``e-box\rq''), and the power and network connections to the observatory.}
\label{fig:block-diagram}
\end{figure*}

SoCal utilizes the same principles as existing, proven solar feeds at other EPRV facilities. Like these instruments, SoCal focuses sunlight through a small (75 mm) lens into an integrating sphere, a hollow sphere internally coated with highly reflective material (Polytetrafluoroethylene, PTFE, aka Teflon). After $\sim$$1000$ reflections within the integrating sphere, a fraction of light rays will eventually land on the tip of a 200~$\mu$m optical fiber that is connected to a port on the side of the sphere. This process spatially scrambles the light from the resolved solar disk and produces a highly homogenized ``point-source-like'' output. The disk-integrated sunlight travels through $\sim$$90$~m of fiber from SoCal on the WMKO roof to the KPF calibration bench in the WMKO basement, where a shutter and beamsplitter allow solar light to be injected into the KPF science (SCI), sky (SKY), and calibration (CAL) fibers, or combinations of them. Figure~\ref{fig:block-diagram} illustrates the full optical path for KPF-SoCal.

\begin{figure}
    \centering
    \includegraphics[width=0.45\textwidth]{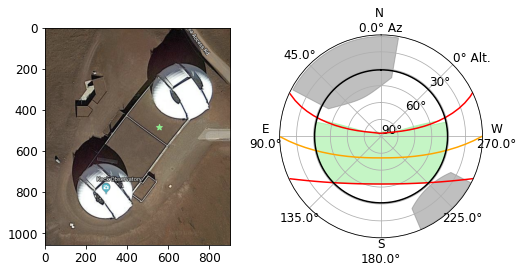}
    \includegraphics[width=0.45\textwidth]{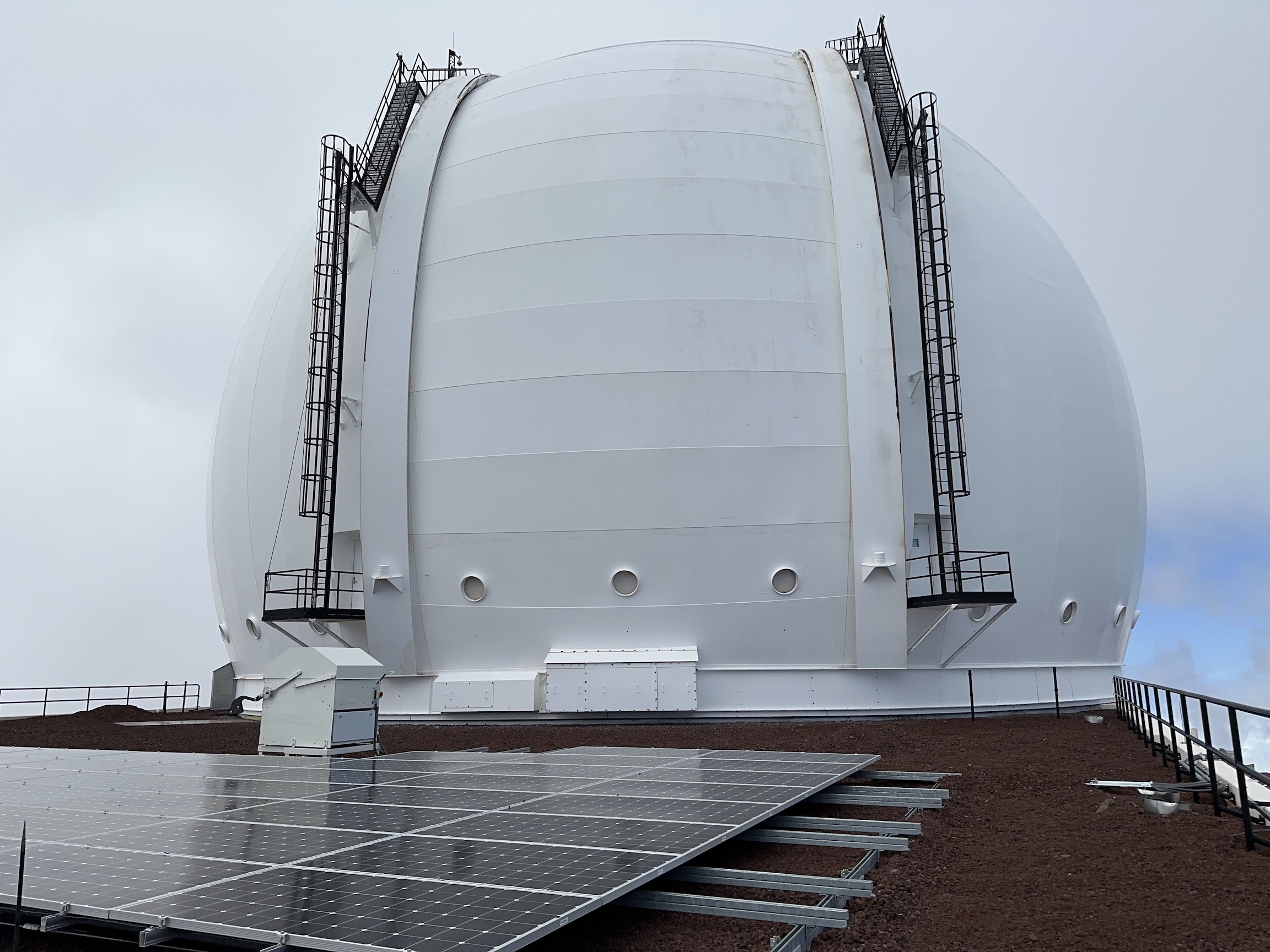}
    \caption{Top: SoCal's location on the WMKO roof (green star in the satellite image on the left), and the shadows of the Keck I and Keck II domes projected on the sky from this location. The path of the Sun is shown for the summer and winter solstices in red and equinoxes in orange. The parts of the sky that the Sun traverses above airmass~$< 2$ (black circle) during a full year are highlighted in green. Bottom: Image showing the SoCal enclosure adjacent to the solar panels, with Keck II in the background.}
    \label{fig:roof-location}
\end{figure}

The main driving design principles for SoCal were 1. enable EPRV-quality stellar activity studies, 2. provide long-term instrumental calibration/tracking, and 3. be robust to the extreme weather environment on the summit of Maunakea. We selected a location on the observatory roof between the Keck I and Keck II domes that would maximize the amount of time during the year when the Sun is observable above airmass~$< 2$ and is not shadowed by either of the Keck domes. Because nearly all of the WMKO roof is tiled with solar panels, SoCal is positioned near Keck II (see Figure~\ref{fig:roof-location}). This location does place SoCal just inside the 30 ft boundary from the Keck II dome where there is a risk of ice sheets sliding off the Keck II dome and falling. Since there were no other locations on the roof with year-round unobstructed access to the Sun, we accepted this small risk.

\subsection{Tracker and Optical System}

\begin{figure}
\centering
\includegraphics[width=0.45\textwidth]{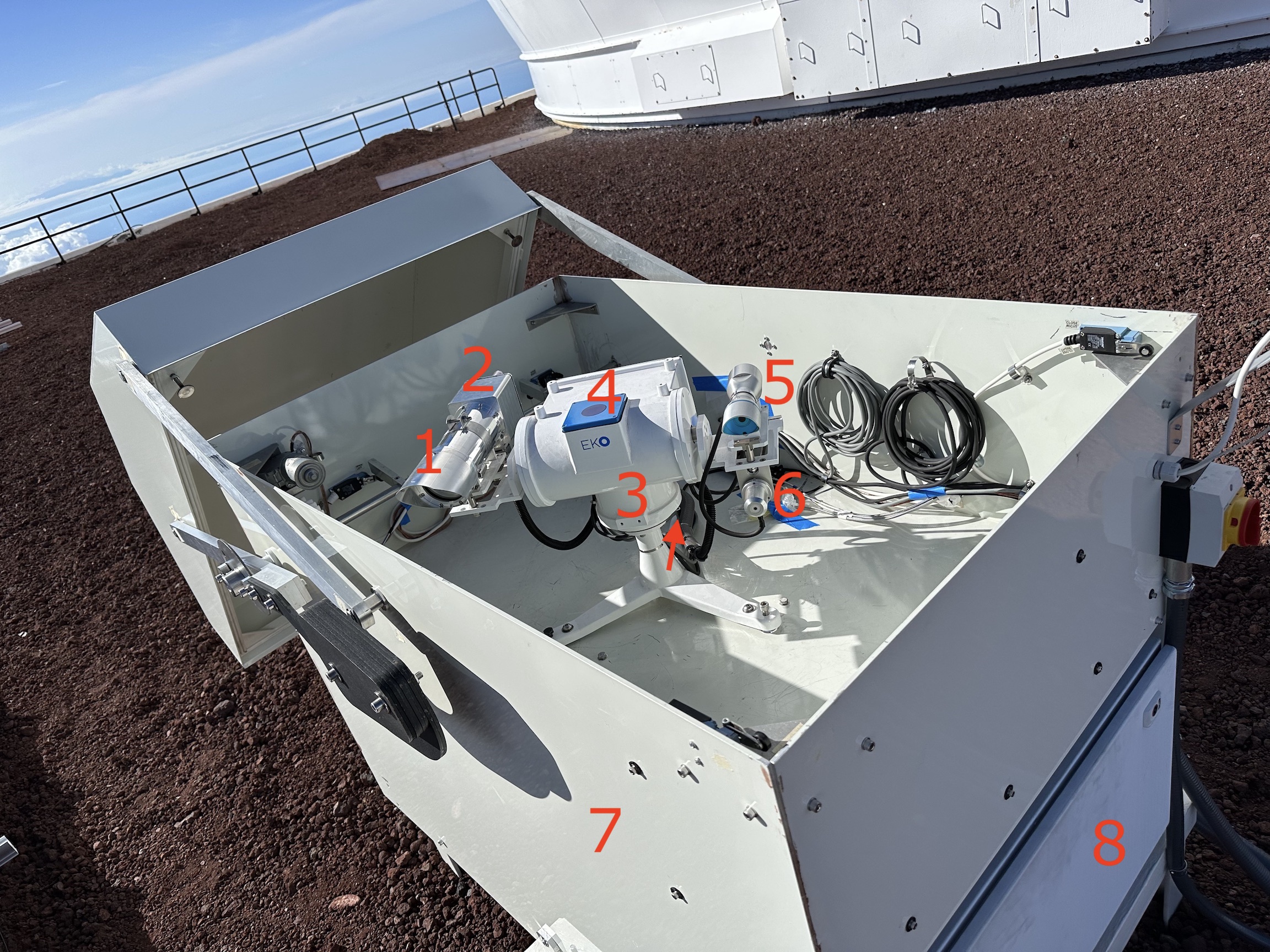}
\caption{The SoCal tracker with mounted optics, inside its enclosure with the lid open. The numbered components are 1. Lens and lens tube, 2. Integrating sphere, 3. EKO Sun Tracker, 4. GPS sensor, 5. Pyrheliometer, 6. Sun-sensor, 7. Enclosure, and 8. E-box. The arrow indicates geographic north; i.e., the tracker is pointed south in this image.}
\label{fig:socal_tracker}
\end{figure}

The optical system of SoCal inherits many design aspects from proven, existing solar feeds at other EPRV facilities, particularly the NEID Solar Feed at the WIYN 3.5~m Telescope at Kitt Peak National Observatory, which largely made use of commercial off-the-shelf (COTS) parts for most components~\citep{Lin2022}. This was especially desirable as we could quickly obtain a working system to test with KPF during the Assembly, Integration, and Testing (AIT) phase of the development of KPF at the Space Sciences Lab (SSL) at UC Berkeley. See Table~2 in \cite{Lin2022} for a list of all of the major components in the sun tracker, pyrheliometer, lens \& lens tube housing, integrating sphere, and shutter mechanism, which are identical for SoCal. The SoCal tracker and optics are shown inside the enclosure in Figure~\ref{fig:socal_tracker}. Here we summarize each briefly.

We purchased the same commercial sun-tracking mount as the NEID Solar Feed, the EKO STR-22G, which has been successfully operated in extreme environments for decades-long experiments (a number of such trackers currently operate at Mauna Loa Observatory). The tracker is an alt-az mount and comes with a built-in quad-pixel ``sun sensor,'' which has provided exceptional active guiding performance for the NEID Solar Feed (see Section~2.4.1 in \citealt{Lin2022}) and also produces helpful telemetry for assessing guiding stability. The tracker is bolted to the upper level of an enclosure with the ``North'' leg of the tripod aligned with geographic north (see Figure~\ref{fig:socal_tracker}). Normally this alignment does not need to be very precise as the sun sensor has a 15$^\circ$ field-of-view and will find and guide on the Sun using an onboard control loop. However, because SoCal is in the tropics, the Sun will often reach elevations of $87^\circ$--$90^\circ$ (see the top panel in Figure~\ref{fig:roof-location}) where the active guiding capability is not possible with this tracker. When the Sun passes near zenith, the azimuth angle rotates a full 180$^\circ$. Because of this, the tracker operates using a predictive calculation above $87^\circ$ by which an onboard GPS sensor uses the device's latitude, longitude, elevation, and current time to adjust the azimuth to the Sun's expected azimuth, while the elevation angle is still adjusted using the sun sensor. As such, precise horizontal leveling and true north alignment are critical for tracking through solar noon. We aligned the tracker to true north by iteratively rotating the tripod by hand, enabling active-guide mode, noting how far the tracker adjusts compared to the predicted position, and then realigning the tripod to minimize the difference between the predicted position and the active-guided position. While the tracker faces south to follow the Sun through the sky during most of the year, for several weeks around the summer solstice the tracker instead must face north. As such, a careful treatment of the cable wrap behind the tracker was needed to accommodate a near $360^\circ$ rotation without snagging.

We also adopted the achromatic lens (3-inch Edmund Optics 88–596-INK), custom aluminum housing, and Thorlabs integrating sphere used in the NEID solar feed \citep{Lin2022}, as these have also been demonstrated to perform well in exposed outdoor environments. \citet{Lin2022} also performed a trade study for several different lens choices and found that this model lens best preserved the size of the Sun in the focal plane across the wide wavelength range of the NEID spectrometer while maintaining sufficient transmission and aperture size. The integrating sphere used is a COTS Thorlabs 2P3 2-inch integrating sphere. The entrance port of the sphere is placed in the approximate focal plane of the lens, centered on the position of the Sun. It is essential to have the image of the Sun formed in free space to avoid overheating the optical components and minimize the risk of vignetting the solar disk.

Like the NEID Solar Feed, SoCal uses the EKO MS-57 pyrheliometer mounted to the secondary arm of the sun tracker to monitor cloud coverage directly in front of the Sun. A pyrheliometer measures the direct normal irradiance (DNI) from the Sun by focusing photons (200--4000~nm) within a narrow range of incident angles ($5^\circ$ field-of-view) onto a blackbody which then radiates to a thermopile. The thermopile converts heat to an output voltage that is proportional to the incident flux, allowing for a simple conversion to \Wmsq{} by multiplying by the factory-calibrated sensitivity (7.717~$\mu$V$/$\Wmsq{} in our case, as listed on the pyrheliometer spec sheet and the device itself). The voltage is automatically converted to irradiance by an EKO MC-20 signal converter, which outputs the resulting data packet in Modbus format. A Lantronix UDS1100-IAP provides a TCP/IP interface for a computer to regularly poll this data once per second.

The shutter assembly on the KPF calibration bench is also a similar design to the NEID Solar Feed shutter assembly. Light from the SoCal delivery fiber is reimaged onto a downstream fiber using a pair of achromats. In the collimated space between the lenses, a Uniblitz shutter provides source selection to the KPF calibration fibers. As SoCal contains two separate optical fibers (one for feeding the KPF science fiber, the other for feeding the dedicated calibration fiber), an identical shutter assembly is used for the second fiber.

\subsection{Optical Fibers \& Path to KPF}\label{sec:fibers}

\begin{figure}
\centering
\includegraphics[width=1.0\columnwidth]{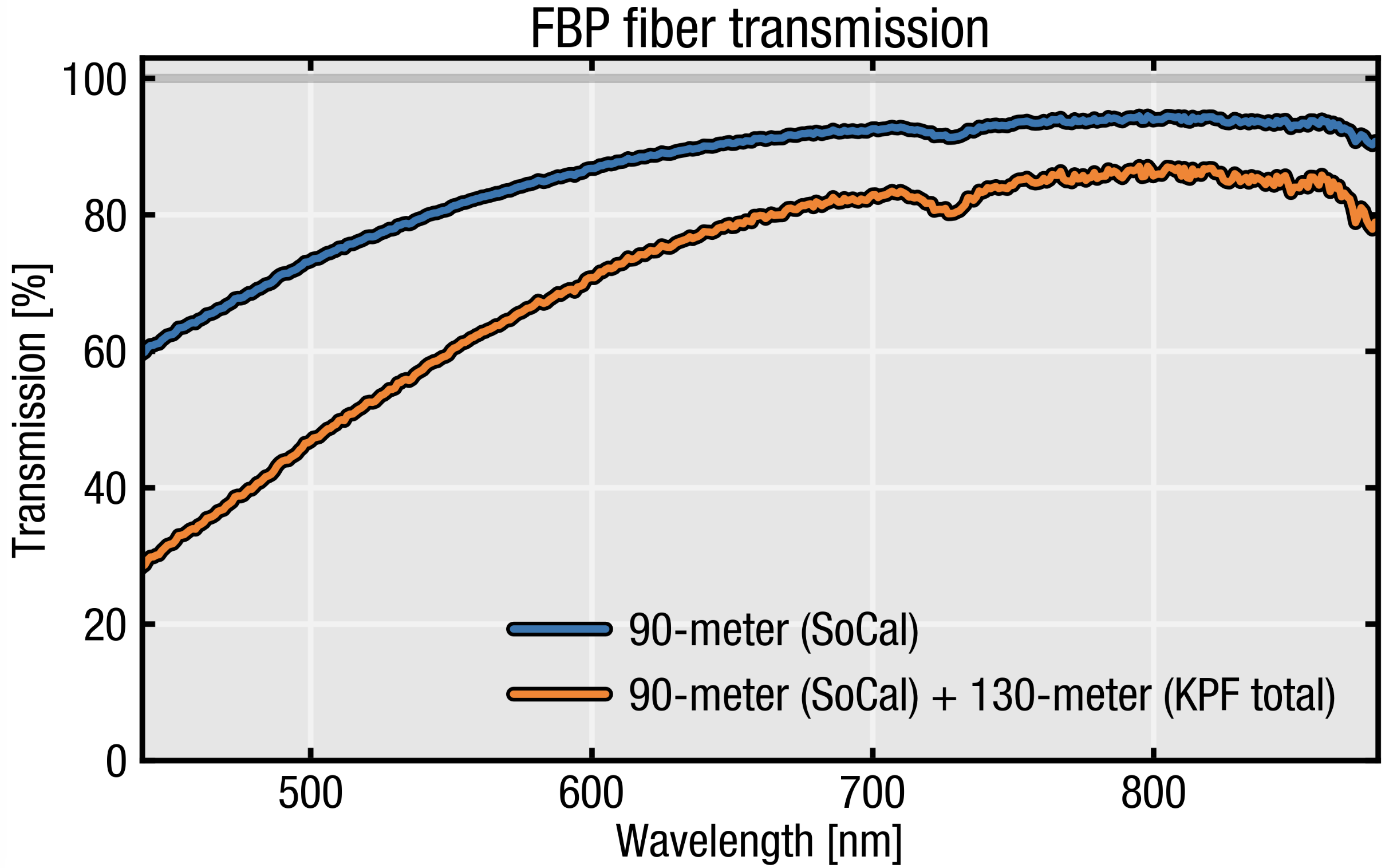}
\caption{Transmission of the 90 meter SoCal fiber run from the Keck Observatory roof to the KPF Calibration Bench in the basement (blue), and the additional fiber run (orange, 130 meters) which includes the KPF calibration fiber (from the Calibration Bench to KPF's Fiber Injection Unit mounted on the Keck I telescope) and science fiber cable (back from the telescope to KPF spectrometer in the observatory basement).}
\label{fig:fiber_throughput}
\end{figure}

The light path and interface between SoCal and the rest of KPF and WMKO are shown in Figure~\ref{fig:block-diagram}. Four separate fiber runs connect SoCal on the roof to the WMKO basement: ``SoCal-SCI,'' ``SoCal-CAL,'' and a spare fiber terminate at the KPF calibration bench, while a separate fiber run for the upcoming HISPEC instrument \citep{HISPEC} is terminated in the basement near the future location of HISPEC. All four fiber runs exit the enclosure, enter a long conduit run along the roof, then enter the building and travel down into the observatory basement for a total length of 90~m.

In the enclosure, a 1-meter COTS Thorlabs $2\times1$ fiber (2$\times$200 ${\mu}$m) fan-out cable plugs into the integrating sphere and splits the collected light into two output fibers, each of which terminates at an FC/PC patch panel at the base of the enclosure. Connected on the other side of the patch panel are the two stainless steel jacketed 90~m fiber runs for SoCal-SCI and SoCal-CAL. The HISPEC fiber is currently capped at both ends, with the output end coiled up near the installation location for HISPEC (also in the WMKO basement). The spare fiber is similarly capped at both ends and serves as a drop-in replacement for any of the other three fibers. 

After traveling 90~m from the roof to the basement, sunlight reaches the KPF calibration bench. Here, the SoCal-SCI fiber injects sunlight into a shutter system. A patch fiber then directs the sunlight into a beam-splitter, where a pair of fibers transport the sunlight from the basement to the Fiber Injection Unit (FIU) on the Nasmyth platform of the Keck~I telescope. Once at the FIU, sunlight follows the same path as starlight collected by the Keck~I telescope. That is, solar photons are injected into the main SCI and SKY fibers in the FIU and transported to KPF via the same optical path as is used for nighttime observations. The non-common path between sunlight and starlight is thus everything before the FIU; the Keck I telescope is used to bring starlight to the FIU while the SoCal optics and roof--to--FIU fiber run does the same for sunlight. In this mode, KPF sees the Sun as a point source just like it would any other star, hence we call this ``Sun-as-a-star'' mode. Likewise it is possible to observe the Sun and a simultaneous calibration source, such as the etalon. The main spectrometer receives four copies of the solar spectrum (the Sky fiber plus the three ``slices'' of the Science fiber) and the Ca~H\&K spectrometer can be simultaneously illuminated.

The SoCal-CAL fiber connects directly to the SoCal port on the KPF calibration source selector, allowing sunlight to be injected into the Cal fiber just like any other calibration source. Hence, this mode is called ``Sun-as-a-calibrant'' mode.

The on-sky performance of SoCal matches predictions made during the planning phase based on estimates of the throughput of the KPF and SoCal systems and the KPF Exposure Time Calculator\footnote{\href{https://github.com/california-Planet-Search/KPF-ETC}{https://github.com/california-Planet-Search/KPF-ETC}}. The system throughput up to the entrance to the fibers in the integrating sphere is  $5.6\times10^{-6}$; that is, the atmosphere, lens, and integrating sphere reduce the amount of sunlight injected into the SoCal fibers by a factor of $\sim 1.8\times10^5$ (the integrating sphere has a reflectivity of $\sim$0.99 and a photon has of order 1000 internal reflections before entering a fiber). The roof--basement--FIU fiber run and numerous optical interfaces along the way introduce another factor of $\sim$40 in flux loss (see Figure~\ref{fig:fiber_throughput} for the throughput contribution from the fibers themselves). Overall, as seen by KPF, the flux of the Sun through SoCal is comparable to the flux from a $V = 1$ magnitude star using the Keck I telescope.

\subsection{Enclosure}\label{sec:enclosure}

To maximize the science productivity (e.g., tracking solar activity over the 11-year solar cycle) and provide long-term instrumental characterization, SoCal needs to operate on a nearly daily basis for many years. To do so, it must survive the extreme weather conditions on Maunakea. Winds in excess of 100 mph (gusts exceeding 150 mph) and significant ice/snow storms are common in the winter months. Weather is generally stable in the summer, although the chance of a tropical storm or (more rarely) a hurricane is ever-present. In November 2022, the eruption of Mauna Loa deposited volcanic ash particulates and ``Pele's hair'' at WMKO, which can damage sensitive optics and equipment. With the additional (though unlikely) risk of ice falling off the Keck~II dome onto SoCal, it was necessary to design and build a protective enclosure for SoCal to weather these natural phenomena.

Other EPRV solar feeds have approached this problem differently. The HARPS-N/LCST and HARPS/HELIOS solar telescopes are each completely enclosed beneath an acrylic dome \citep{Phillips2016}. This has the benefit of no moving parts but requires heat management and introduces the potential for aberrations, as scratches or imperfections on the dome could distort the solar image and produce spurious (possibly chromatic) RV shifts. In fact, accumulated dust on the HELIOS dome is likely responsible for observed oscillations in some of the HARPS solar RVs \citep{Zhao2023}. Meanwhile, the GIARPS/LOCNES solar telescope lives inside a small aluminum box with a motorized lid~\citep{Claudi2018}. In contrast, the NEID solar feed does not use an enclosure of any kind; \citet{Lin2022} instead opted for highly ruggedized components for the entire system, which is mounted in the open on the roof of the WIYN control room building. 

As the sun tracker, pyrheliometer, lens, lens tube assembly, and associated cables are all similar to those in to the NEID solar feed, which itself has weathered monsoons and survived being fully encased in ice, we have confidence that SoCal can likewise withstand significant weather events. In fact, a number of EKO STR-22G Sun Trackers operate enclosure-less at nearby Mauna Loa Observatory, which experiences similar weather. However, due to the high frequency of hurricane-force winds, which could uplift cinder and impact SoCal, and the chance of ice fall, we opted for a rugged, highly-weatherproofed motorized enclosure to protect the sun tracker and optics. Additionally, keeping SoCal covered when not in use reduces UV degradation, extending the lifespan of various components.

\begin{figure}
\centering
\includegraphics[width=0.45\textwidth]{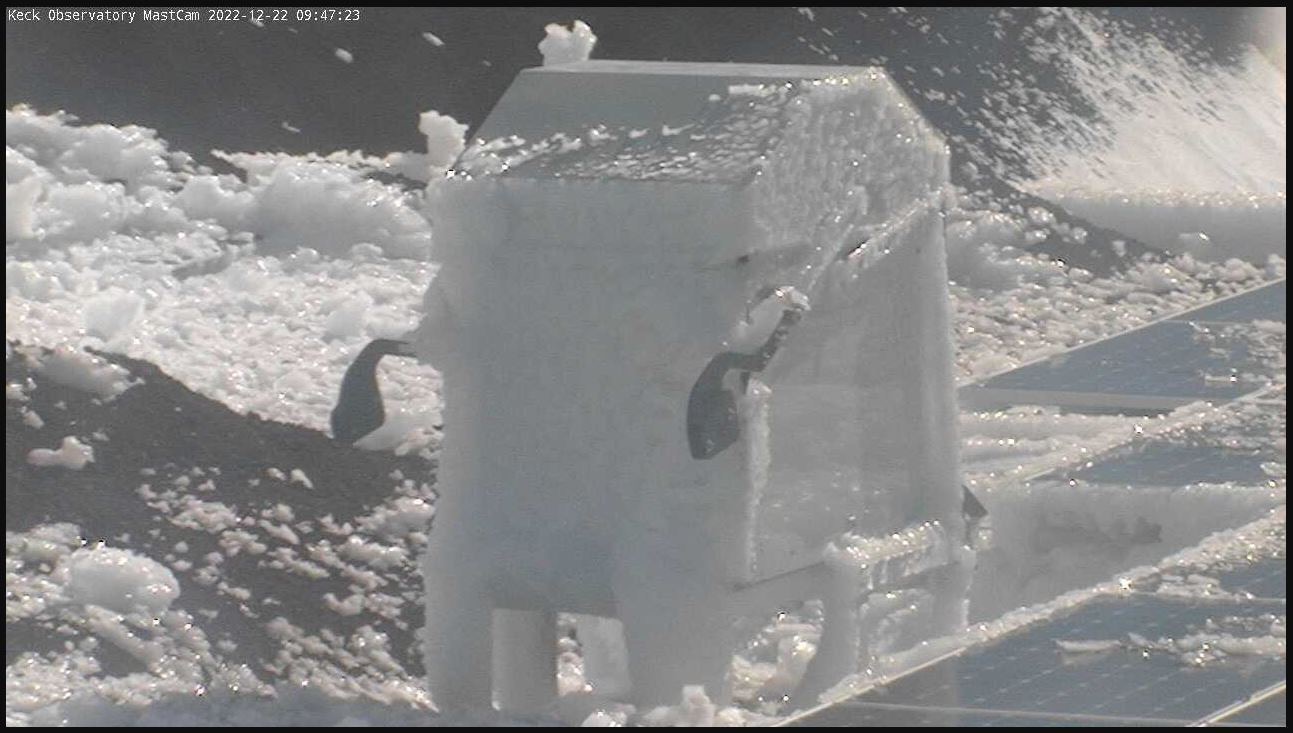}
\caption{Webcam image of the SoCal enclosure frozen in a block of ice after a winter storm in December 2022.}
\label{fig:socal-storm}
\end{figure}

We opted for a proven solution to shield SoCal from the elements when necessary; the clamshell-style design of our enclosure is the same as that for the Hungarian-made Automated Telescope Network (HATNet), also on Maunakea \citep{Bakos2002, Bakos2004}, and was custom-built for our purposes by the same manufacturer, Fornax Mounts\footnote{\href{https://fornaxmounts.com/}{https://fornaxmounts.com/}}. The SoCal enclosure is visible in Figures~\ref{fig:roof-location}, \ref{fig:socal_tracker}, and \ref{fig:socal-storm}. The enclosure control electronics, called ``Dome Guard,'' are controlled by a Raspberry Pi single-board computer. A sensor monitors the current sent to the dome motor and cuts off the power if the measured current exceeds a user-specified threshold. This prevents the lid from opening into an obstruction (e.g., a snowbank) and straining the worm gear/cogwheel.

The SoCal enclosure is kept in place by burying attached steel plates under the layer of 0.5~m thick volcanic cinder that covers the roof of the building connecting Keck I and II.  This approach was adopted so that it wouldn't be necessary to puncture the water-tight membrane on the roof.  The enclosure frame is welded to six legs which were bolted to the steel plates. The frame and foundation were designed by M3 Engineering \& Technology, who designed an analogous ballasted mounting scheme for WMKO's solar panel array\footnote{\href{https://m3eng.com/portfolio/w-m-keck-observatory-solar-panel-supports/}{https://m3eng.com/}}. The weight and area of the steel plates were determined by considering the wind-loading of the enclosure to ensure that the combined weight of the enclosure ($\sim$700 lbs), steel frame and foundation ($\sim$700 lbs), and backfilled cinder would be sufficient to withstand winds up to 200~mph (3~sec gust). Mechanical latches were also installed to securely hold the lid in its closed position in anticipation of strong winds.

The enclosure was installed on December 16, 2022, with a partially assembled sun tracker inside. Two days later the summit experienced an extreme winter storm with sustained winds in excess of 100 mph and severe snow/ice (see Figure~\ref{fig:socal-storm}). The severe winter weather continued for roughly four months before we were able to return and complete the installation. The enclosure successfully protected the sun tracker inside; only a slight dusting of volcanic cinder was found in the interior, which was wiped away with a cloth.

\subsection{Electronics}\label{sec:electronics}

The control electronics for SoCal are stored in a dedicated weatherproof box (the ``e-box'') stowed in the lower level of the enclosure (visible in the lower-right corner of Figure~\ref{fig:socal_tracker}). Most cabling is internal; the only external electrical cables are a 120~V/15~A AC power cable and a pair of cat-6 ethernet cables (main and spare). Power is grounded in the same manner as other rooftop devices and the cat-6 cables have surge suppressors in series to protect against lightning strikes. Two power supplies (24~V DC and 12~V DC) inside the e-box supply power to all electronics devices. The 24~V devices (including the tracker and enclosure) have backup power from an uninterruptible power supply (UPS). In the event of a power failure, the enclosure control logic detects the switch to the UPS and automatically commands the lid to close using backup power.

Our primary selection criterion for the various electronics devices was the ability to operate in a wide operating temperature range. While the ambient temperature on the summit is generally stable (T~$\approx 0 \pm 10^\circ$~C), we set a conservative requirement of $-30^\circ$~C to 60$^\circ$~C operating temperature as temperatures can be more extreme inside the sealed e-box in shaded or direct sunlight conditions. The temperature inside the e-box, inside the enclosure, and outside the enclosure are each monitored using a dedicated temperature probe.

We also required TCP/IP interfaces for each device to integrate with the KPF and WMKO facility networks. The sun tracker communicates using RS232, so a Lantronix UDS2100 provides direct control over TCP/IP. The output voltage from the pyrheliometer is converted into Modbus protocol using an EKO MC-20 signal converter, and a Lantronix UDS1100-IAP provides the TCP/IP interface. The enclosure is fully operable over WebSocket so it is directly controlled without an additional device server. Since the enclosure has its own computer, it monitors connections to its IP address and automatically triggers the lid to close should it lose connection to the KPF server.

\subsection{Control Software}\label{sec:software}

The SoCal system consists of three main devices: the sun tracker, the pyrheliometer, and the enclosure. We communicate with each device using Python functions. We use the \texttt{socket} module to send RS232 commands to the sun tracker, \texttt{pymodbus} to poll data from the pyrheliometer, and \texttt{websockets} to communicate with the Dome Guard Raspberry Pi single-board computer. 

\begin{figure*}
    \centering
    \includegraphics[width=0.95\textwidth]{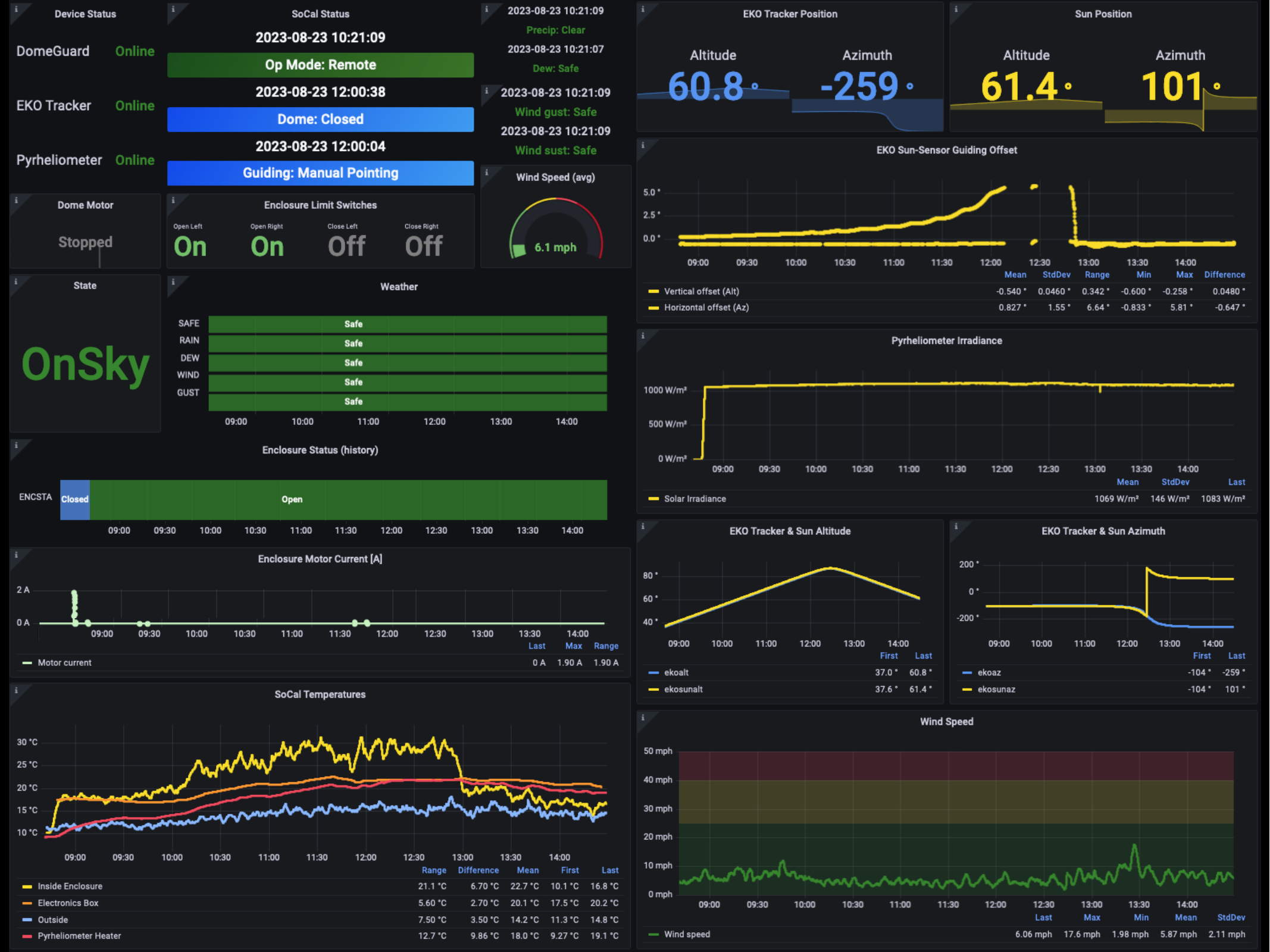}
    \caption{Screenshot of the Grafana web page displaying the SoCal telemetry for an example day. Grafana is a web-based interactive visualization software for displaying and plotting values from a database. In this example near the summer solstice, the tracker performed a large slew in azimuth through solar noon at near 90$^\circ$ elevations. The sun sensor guiding offset is plotted in the upper right-hand corner (box titled ``EKO Sun-Sensor Guiding Offset''). Near solar noon the sun tracker switches to predictive guiding mode, hence the gap in recorded guider offsets. The elevation offset is stable at $\sim 0.5^\circ \pm 0.05^\circ$ all day. While the azimuth offset increases near zenith, the actual angular separation between the predicted Sun location and the sun tracker's position is never more than $\sim 0.5^\circ$. Other panels display weather information and the status of subsystems.}
    \label{fig:grafana}
\end{figure*}

For operations at WMKO, these Python functions are wrapped into the Keck Task Library \citep[KTL;][]{KTLkeyword, KTL, Deich2014} keyword framework using \texttt{KTLPython}. This allows users to interface with SoCal using the same syntax as is used for other Keck instruments. Telemetry is stored using a set of KTL keywords.  The history of these keywords (e.g. tracker altitude, sun sensor offset, enclosure open/close state, temperatures) is stored in a database on WMKO servers which can be queried to determine the current or past state of SoCal. One way to visualize this information is through a Grafana web page (Figure~\ref{fig:grafana}).

\section{Operations}\label{sec:operations}

\begin{figure*}
    \centering
    \includegraphics[width=0.95\textwidth]{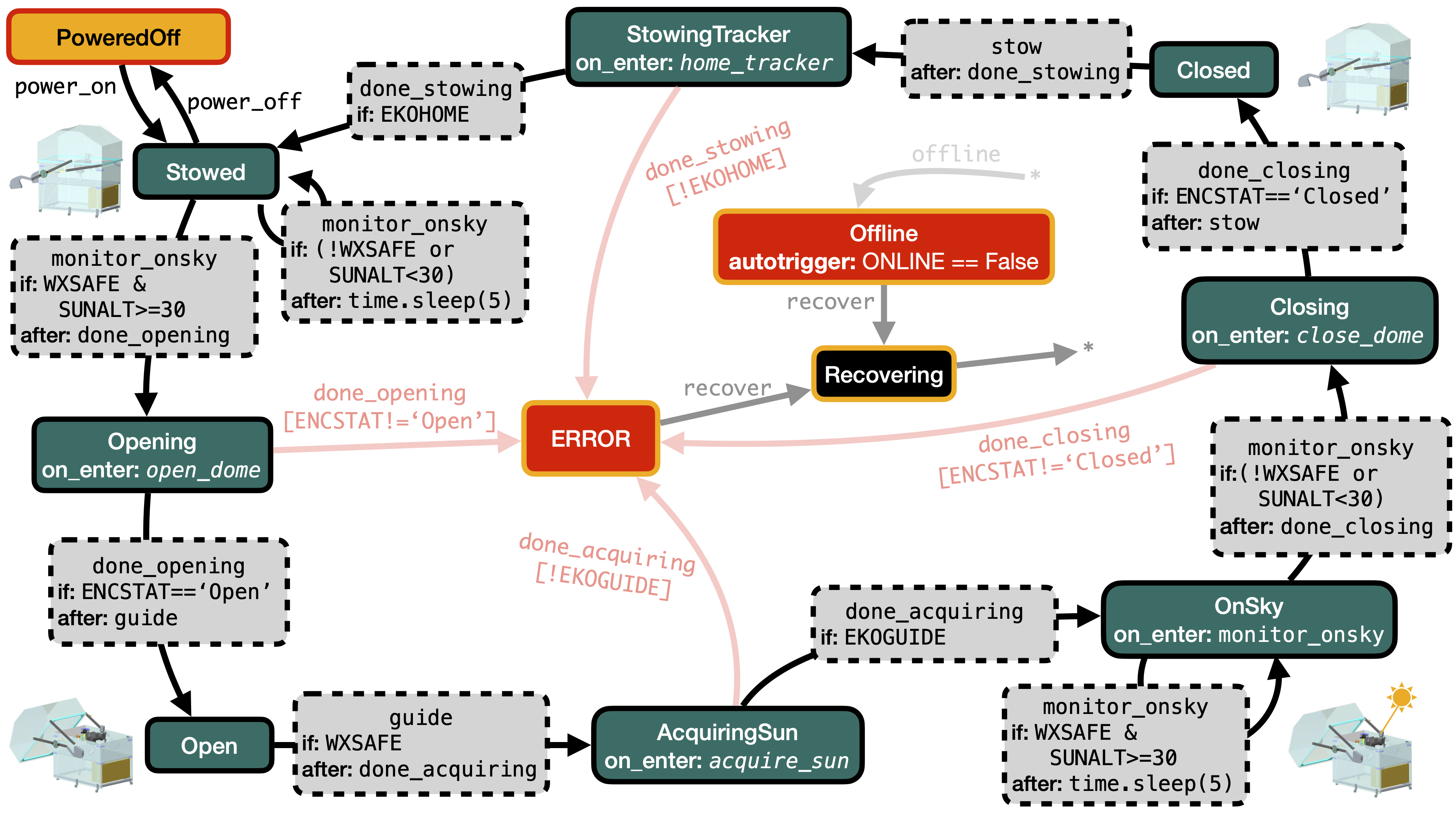}
    \caption{State machine logic flowchart defining the automation loop for SoCal operations. Nominal operations begin in the upper left and flow counter-clockwise. First, the enclosure opens and the tracker acquires the Sun. As the tracker guides on the Sun, the KPF spectrometer records solar spectra. At the end of the day, the dome closes, and the tracker is stowed (see Section~\ref{sec:state-machine} for more details). Green boxes with solid borders represent the states of the system, while grey boxes with dashed borders represent transitions between states. The special states ``Offline'' and ``ERROR'' are visualized with red boxes, while ``Recovering'' is colored black. The state to which a named transition moves to may depend on a conditional, which is printed as an \texttt{if} statement. While most states execute a function upon entering (\texttt{on\_enter}), the \texttt{Open} and \texttt{Closed} states generally immediately transition to the ensuing state due to the \texttt{after} function of the transition used to enter those states.}
    \label{fig:state-machine}
\end{figure*}

\subsection{Daily Schedule}\label{sec:daily-operations}

The daily calibration schedule for KPF consists of a set of morning and evening calibrations with the fiber illuminated by Thorium-Argon (ThAr) and Uranium-Neon (UNe) hollow cathode lamps, a broadband laser frequency comb (LFC), a stabilized Fabry-Perot etalon, and a quartz flat lamp, plus dark and bias frames. These automated calibration sequences are scheduled at fixed times, with morning calibrations ending around 08:42 HST (Sun at 30--40$^\circ$ elevation, or airmass 2--1.6) and evening calibrations beginning at 3:00 HST (Sun at 30--55$^\circ$, or airmass 2--1.2). This leaves roughly $\sim$6 hours of available daytime for SoCal, year-round. Currently, the time from noon HST to 3:00 HST is used to collect continuous stacks of flat-field frames, so thus far SoCal has operated in the morning hours from 08:45 -- noon HST. We preferred morning over evening as the former overlaps with solar observations in Arizona by both NEID and EXPRES. Long term, we are exploring scheduling flat-field calibrations during off-sky nighttime hours to free up the full daytime for SoCal. An additional $\sim$2 hours of daytime in the summer months may be obtained by dynamically scheduling calibrations according to sunrise/sunset times. However, this would result in the morning/evening calibration sequences occurring at different relative times to the fixed liquid nitrogen fill schedule ($\sim$11:00 HST) throughout the year.

Following the morning calibration sequence, the SoCal observing script initiates operations. This script first checks that SoCal is in the \texttt{OnSky} state (i.e., the sun tracker is guiding on the Sun above $30^\circ$ elevation, see Section~\ref{sec:state-machine} for more details). If so, the script configures the FIU to select the SCI and SKY calibration fibers, configures KPF to use the green and red CCDs as well as the Ca H\&K spectrometer, configures and activates the exposure meter, runs the agitator, configures the shutters, and finally turns and directs etalon light into the simultaneous calibration fiber. Then, as long as SoCal remains in the \texttt{OnSky} state, repeated exposures are taken. The autonomous loop regularly monitors (every 5 seconds) weather keywords from the observatory's meteorological system (sustained wind speed, wind gust speed, dew point, and precipitation) and if any become ``unsafe,'' or if the Sun sets, SoCal exits \texttt{OnSky} and exposures terminate. The observing script is re-executed if SoCal re-enters \texttt{OnSky} (e.g., if the weather becomes ``safe'' again) up until the evening calibration sequence is scheduled to begin. We adopted the same ``safe/unsafe'' conditions used for general operations at WMKO, which correspond to a dewpoint temperature within 0.2 C of ambient temperature, wind gusts over 45 mph, and/or sustained wind speeds over 30 mph. We have also noticed that at wind speeds near 30 mph, the enclosure lid visibly bounces up and down as its concave shape in the open position acts as a sail. Thus, keeping the enclosure closed in strong winds reduces strain on the mechanical components.

KPF exposures with SoCal are taken with a fixed exposure time of 5~sec (see discussion in Section~\ref{sec:results}). KPF has two readout modes, ``standard'' and ``fast readout.'' The fast readout mode is primarily used for high-cadence asteroseismology during nighttime operations. Initial SoCal operations during commissioning were primarily in standard mode, which originally had a 55~sec readout time but has since been reduced to 49~sec. For comparison,  the cadence of the NEID solar feed is 83~sec (55~s exposures + 28~sec readout), which is similar to SoCal's cadence in standard read mode. During most mornings with the current operating scheme, KPF records around 200 solar spectra in standard readout mode. Long-term post-commissioning SoCal operations are expected to utilize the fast readout mode (15~sec) to produce daily time series of solar spectra with $< 30$~sec cadence. In this mode, SoCal will accumulate $\sim$1000 spectra per 6~hr day. KPF has been tested with SoCal in fast readout mode on a few days, including a single full 6~hr day during which 1041 spectra were acquired.

\subsection{Autonomous Loop}\label{sec:state-machine}

SoCal is autonomously controlled using state-machine logic that transitions the system between defined states. A state machine works by defining a number of known ``states'' which correspond to different configurations of the various devices in the system. ``Transitions'' define how one state moves to another. Pre-condition and post-condition functions can be attached to each state and transition so that they are executed before or after a transition, or upon entering or exiting a defined state.

The autonomous loop, which we implement using \texttt{pytransitions} \citep{pytransitions}, is shown graphically in Figure~\ref{fig:state-machine} and is as follows. Beginning in the \texttt{Stowed} state, with the enclosure closed and the sun tracker pointed at ``home'' (due south at zero elevation), the \texttt{monitor\_onsky} transition is called. This transition checks if the Sun is above the horizon and if all weather keywords report ``safe.'' If false, the transition returns to the \texttt{Stowed} state, waits five seconds, and attempts to transition again. If true, the state machine transitions to \texttt{Opening}, and the enclosure is commanded to open. After opening, the sun tracker is set to active guiding mode. Upon acquiring the Sun (defined by the sun sensor guiding offset falling below $1^\circ$), SoCal enters the \texttt{OnSky} state. Five seconds later the software checks if the weather keywords are all ``safe'' and that the sun altitude is still $> 30^\circ$. If both are true, the state machine transitions to \texttt{OnSky}; in this case, there is no state change, and thus this check every 5~seconds continues. If one of the two conditions fails, then the state machine transitions to \texttt{Closing}, triggering the enclosure to close. Once closed, the sun tracker is commanded to move to its home position. SoCal then reenters the \texttt{Stowed} state, and the whole process starts over.

Three special states exist for gracefully catching errors and recovering without human intervention. The \texttt{Offline} state, which the state machine can transition to from any other state, occurs automatically if a regular ping to any of the SoCal devices fails. The state machine will hold in this state until all devices become ping-able again, at which point the state machine transitions to the \texttt{Recovery} state. The \texttt{ERROR} state is automatically transitioned to if an exception is caught during any of the \texttt{before}/\texttt{after}/\texttt{on\_enter}/\texttt{on\_exit} functions. Similarly, upon entering \texttt{ERROR}, the state machine will attempt a transition to \texttt{Recovery}.

Upon entering \texttt{Recovery}, the code evaluates the status of each SoCal device by requesting the relevant telemetry. If the telemetry is consistent with the last known state, then the state machine transitions back without executing any of the associated \texttt{before}/\texttt{after}/\texttt{on\_enter}/\texttt{on\_exit} functions. Otherwise, the code executes the relevant device commands to put the devices back in the correct configurations to be consistent with the last known state and then transitions to that state. If this too fails, then the state machine remains stuck in the \texttt{ERROR} state. After a timeout the enclosure is commanded to close and an email and Slack message are sent to relevant personnel. The enclosure also has its own hard-wired fail-safes that automatically close the enclosure in the event of a power outage (the UPS provides backup power) or if the enclosure becomes unreachable (flagged by regular pings between the enclosure and KPF computers).

\section{Data Reduction}\label{sec:drp}

\subsection{KPF Data Reduction Pipeline}\label{sec:kpfpipe}
SoCal spectra follow essentially the same data reduction steps as stellar spectra gathered using the Keck I telescope. Each of the three primary KPF science slices is independently extracted and reduced using the standard KPF DRP\footnote{\href{https://github.com/Keck-DataReductionPipelines/KPF-Pipeline/}{https://github.com/Keck-DataReductionPipelines/KPF-Pipeline/}}. RVs are computed using the cross-correlation (CCF) technique, using a weighted numerical stellar mask based on spectral type \citep[e.g.]{Pepe2002, Baranne1996}, for each SCI slice and for each CCD (green and red) independently. The KPF DRP currently uses the public release of the ESPRESSO cross-correlation masks; the G2 mask is used for the solar spectra. The KPF DRP produces three main data products in the form of .fits files: ``Level 0'' (L0) files contain the raw 2D images from the green and red CCDs, ``Level 1'' (L1) files contain the extracted 1D spectra for each fiber trace (three slices for SCI, one for SKY, one for CAL), and ``Level 2'' (L2) files contain the RVs in the green and red channels (averaging over the three slices). We further combine the green and red RVs into a single RV using an unweighted mean.

The main step that requires special treatment for solar data is the barycentric correction \citep{Wright2014}. Using \texttt{barycorrpy} \citep{barycorrpy}, the doppler shift due to the barycentric motion of the Sun due to the Solar System planets as well as the motion of the observatory along the line of sight is removed \citep{Wright2020}. This is accomplished by multiplying the wavelength solution of the CCF mask by $1 / (1 + v_b/c)$, where $v_b$ is the output of \texttt{barycorrpy.get\_BC\_vel} using \texttt{SolSystemTarget=`Sun'} and \texttt{predictive=True}. We also compute and report heliocentric Julian dates (\hjd) using \texttt{barycorrpy.utc\_tdb}, as opposed to barycentric Julian dates (\bjd) computed for stars. Consequently, the final KPF solar RVs are in the rest frame of the Sun. Thus, any observed variability must be due to solar activity, instrumental noise, or atmospheric/resolved-disk effects.

The current KPF DRP implementation does not correct for differential extinction (as described in \citealt{Davies2014} and \citealt{CollierCameron2019}). The L2 files provide information about instrument drift by reporting the RV of the simultaneous calibration (etalon) spectra; these RVs can be subtracted from the solar RVs to correct for the drift.  However, during the first few months of SoCal operations, the etalon has not been consistently available with high enough flux to enable such an RV drift correction due to the etalon illumination source (an NKT supercontinuum laser) degrading over time. For observations since July 31, 2023, the extracted green etalon RVs are too noisy to be useful for a simultaneous drift correction. However, this limitation is expected to be short-lived as a replacement supercontinuum source will be installed in the very near future. With anticipated developments of the KPF DRP, a global drift model for the instrument will be constructed each day based on standalone and simultaneous calibrations taken throughout the day. This model could then be subtracted from the measured RVs for higher-precision measurements.

The L0--L2 solar data are publicly available on the Keck Observatory Archive\footnote{\href{http://koa.ipac.caltech.edu/cgi-bin/KOA/nph-KOAlogin}{http://koa.ipac.caltech.edu/cgi-bin/KOA/nph-KOAlogin}} (KOA) by querying KPF data for \texttt{TARGNAME == `Sun'}, or by using the \texttt{PyKOA} API\footnote{\href{https://koa.ipac.caltech.edu/UserGuide/PyKOA/PyKOA.html}{https://koa.ipac.caltech.edu/UserGuide/PyKOA/PyKOA.html}}. SoCal data are categorized as calibration data and are therefore available for public use within a day of being collected. We expect to add a queryable, downloadable table of SoCal RVs with telemetry and quality control metrics using the pyrheliometer (see Section~\ref{sec:pyr}), as well as the raw irradiance time series for each day that SoCal was active. The irradiance measured during each exposure is also saved as an extension in the L0 file.

\subsection{Quality Control}\label{sec:pyr}

The largest variability in the solar RVs is caused by uneven throughput across the resolved stellar disk. While the integrating sphere spatially averages over the stellar disk to sufficient homogeneity, external factors such as clouds or objects on the horizon can obscure some or all of the solar disk, which breaks the symmetry of the solar rotational velocity profile and creates large time-variable RV shifts up to $v\sin i_\odot \sim 3$~km~s$^{-1}$. Rather than create a quality flag based on the observed solar spectrum or RV, we used the pyrheliometer irradiance time series to identify observations that are contaminated by clouds or other obscurations. If the Sun is partially or completely obscured, the measured irradiance time series from the pyrheliometer shows a drop in flux. As a cloud moves across the solar disk, the irradiance time series can also show erratic variability. Conversely, clear-sky conditions produce a stable, slowly varying irradiance curve that peaks at solar noon.

\begin{figure*}
\centering
\includegraphics[width=0.95\textwidth]{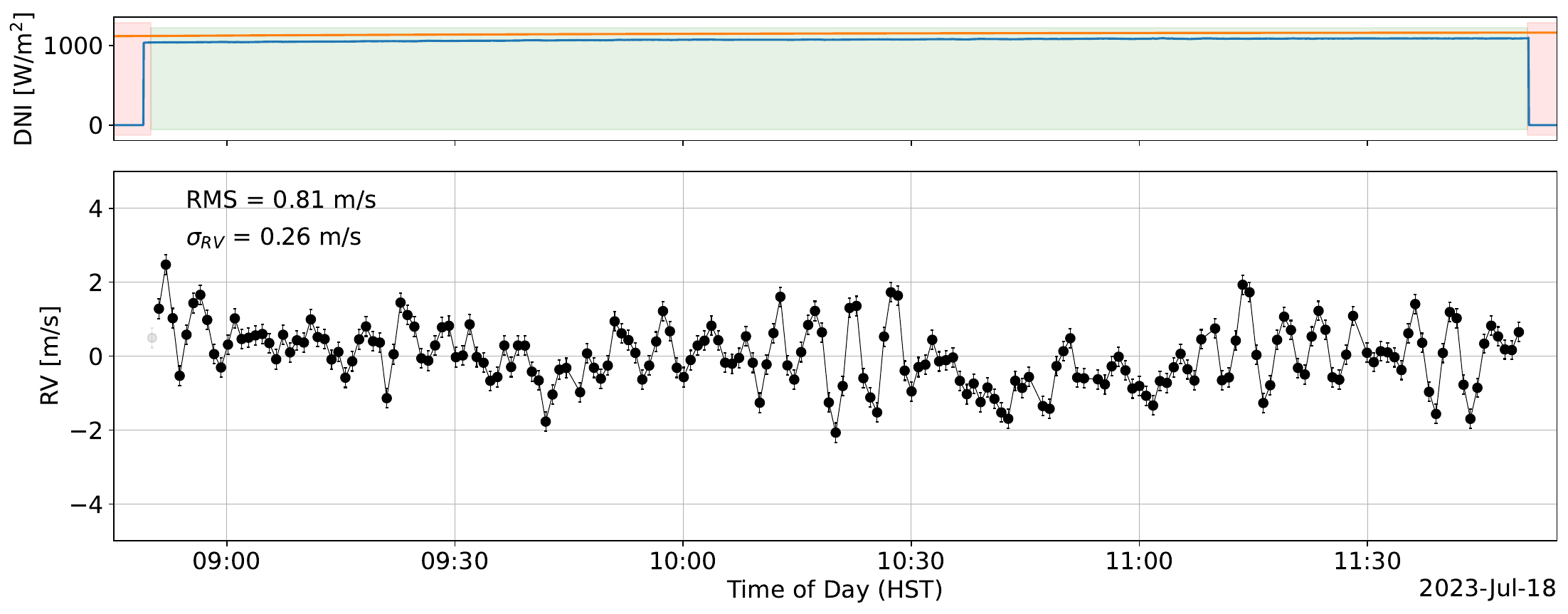}
\includegraphics[width=0.95\textwidth]{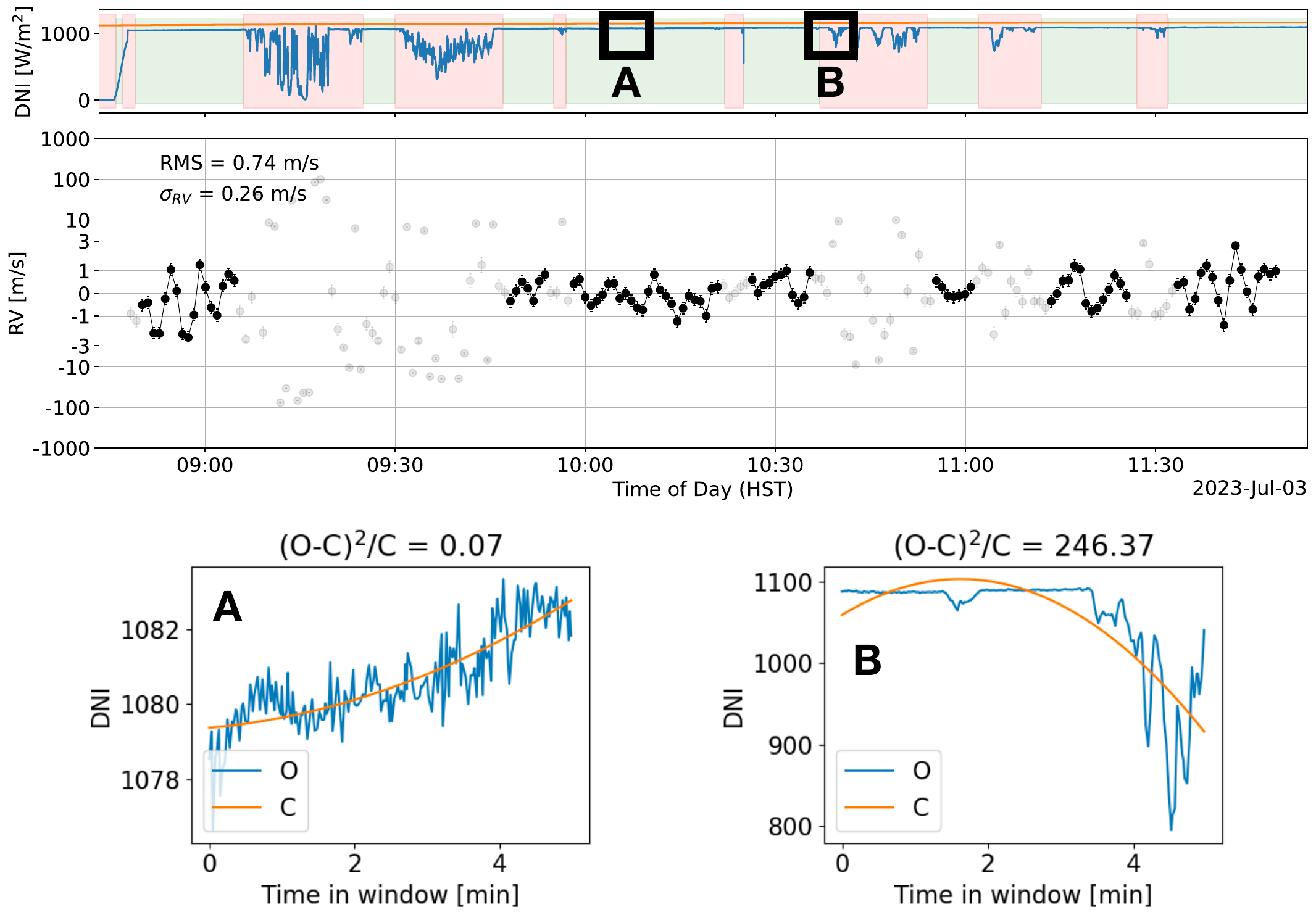}
\caption{Example of a clear-sky day (top) and a day with sporadic clouds (bottom). The upper subplot shows the irradiance time series (blue) relative to the theoretical model (orange) computed using \texttt{pvlib} \citep{pvlib}, with clear times highlighted in green and non-clear times in red as identified by the clearness index defined in Section~\ref{sec:pyr}. The lower subplot plots the RVs during the same time frame. Note the vertical axis scale for the RVs on the cloudy day; RVs observed through clouds show a wide range of sporadic variations from a few to hundreds of m~s$^{-1}$. The $\sim 5.5$~minute p-mode oscillations are clearly seen in the clear-sky RVs; connecting lines are drawn to help guide the eye. Faded points are RVs masked according to the clearness criteria described in Section~\ref{sec:pyr}. The zoom-ins at A and B in the cloudy example show the polynomial fit and resulting clearness index for a reference clear and cloudy window.}
\label{fig:clearness}
\end{figure*}

We devised an algorithm to assess cloud coverage using the irradiance time series. For each day SoCal operates, the irradiance time series is divided into 5~min duration windows. This window size is adjustable, but 5~min was found to be both long enough to include enough data points for a reliable calculation as well as short enough to capture the fast timescale nature of cloud coverage. For each window, the algorithm fits a second-degree polynomial. The ``clearness index'' is the square of the residuals, dividing by the polynomial model, summed over the 5~min window. Essentially, this is a $\chi^2$ test. With clear skies, the polynomial model is a good fit to the stable, slowly varying irradiance, and so the clearness index is low. If clouds are present, the large changes in irradiance produce a high clearness index. We found that setting a threshold of $< 2$ for the clearness index in 5~min windows effectively selects only the clearest portions of the day. Adding a secondary criterion that the observed irradiance be $> 100$~\Wmsq{} eliminates cases where an obstruction causes a decrease in the measured irradiance to near 0~\Wmsq, which would pass the $\chi^2$ test should the zero flux be maintained for the duration of the window.  For finer time resolution and increased robustness at the bin edges, we repeat this calculation five times, each time shifting the windows by one minute. The clearness index at a given timestamp is then the minimum of the values computed from the shifted windows which include that timestamp. Figure~\ref{fig:clearness} shows example clear and cloudy days with times that pass this clearness threshold highlighted in green. Conveniently, the clearness index does not depend on a theoretical model for the irradiance, only the measured time series, and is fast to compute. Figure~\ref{fig:clearness} also shows the corresponding RVs, which are masked (faded points) if the clearness threshold fails or if there is not at least three minutes of clear-sky time. An additional buffer of one minute is masked at any clear/not clear boundary.

Applying this filter to the full set of SoCal observations discards $\sim$16\% of all RVs. We visually inspected the corresponding plot in Figure~\ref{fig:clearness} for each day to ensure that data affected by clouds were being correctly identified.

\section{First Results}\label{sec:results}

We completed the installation of SoCal at WMKO and achieved ``first light'' on April 25, 2023. Initial data were collected on a few clear days in May under manual control while the control software was being finalized and hardware issues that disabled remote operation of the enclosure were resolved. Beginning on June 5, 2023, SoCal and KPF have observed the Sun nearly every ``safe weather'' day (which may or may not be cloudy) as described in Section~\ref{sec:daily-operations}, with occasional shutdowns for testing of other KPF subsystems. SoCal was intended to both assist with KPF commissioning tasks and collect useful data for studying stellar activity. Here we discuss the first results from these activities.

\subsection{Doppler Performance}\label{sec:kpf-validation}

During commissioning, SoCal data were used to validate the Doppler performance of KPF, identify instrumental problems, and provide an additional calibration source and benchmark for the DRP. To validate Doppler performance we have accumulated over 19,000 solar spectra using the standard and fast readout modes over a few to six hours per day during 111 calendar days spanning 4.5 months. The solar spectra were reduced using the KPF DRP as described in Section~\ref{sec:kpfpipe}. Observations taken in cloudy conditions were removed using the ``clearness index'' presented in Section~\ref{sec:pyr}. The KPF DRP is being actively refined and currently works best over short time periods, hence in this work we only scrutinize KPF's performance on intra-day timescales. Future work will probe KPF's Doppler performance on timescales of weeks to months.

In a 5~sec exposure, the extracted 1-D KPF spectra have a peak signal-to-noise (SNR) of $\sim$450 in the green channel ($\sim$550~nm) and $\sim$800 in the red channel ($\sim$750~nm), per SCI trace (the large differences come from the significantly worse throughput at bluer wavelengths from the long fiber run, see Figure~\ref{fig:fiber_throughput}). Combining the three SCI traces yields SNR $\sim$800 in green and $\sim$1400 in red. For reference, nonlinearity in the response of the KPF CCDs is expected to set in for SNR $\gtrsim$1500 in a single trace, or $\sim$2600 combined (true saturation at 1900 and 3300 respectively). Combining the measured green and red RVs yields a photon-limited precision of around $28$~\cms for a given 5~\textit{sec} exposure. Since we also expose the SKY fiber to sunlight, in theory we can gain an additional $\sim$$\sqrt{4/3}$ increase in SNR by combining SKY with the three SCI traces; However, this is currently untested. 

\begin{figure*}
      \centering
    \includegraphics[height=0.25\textheight]{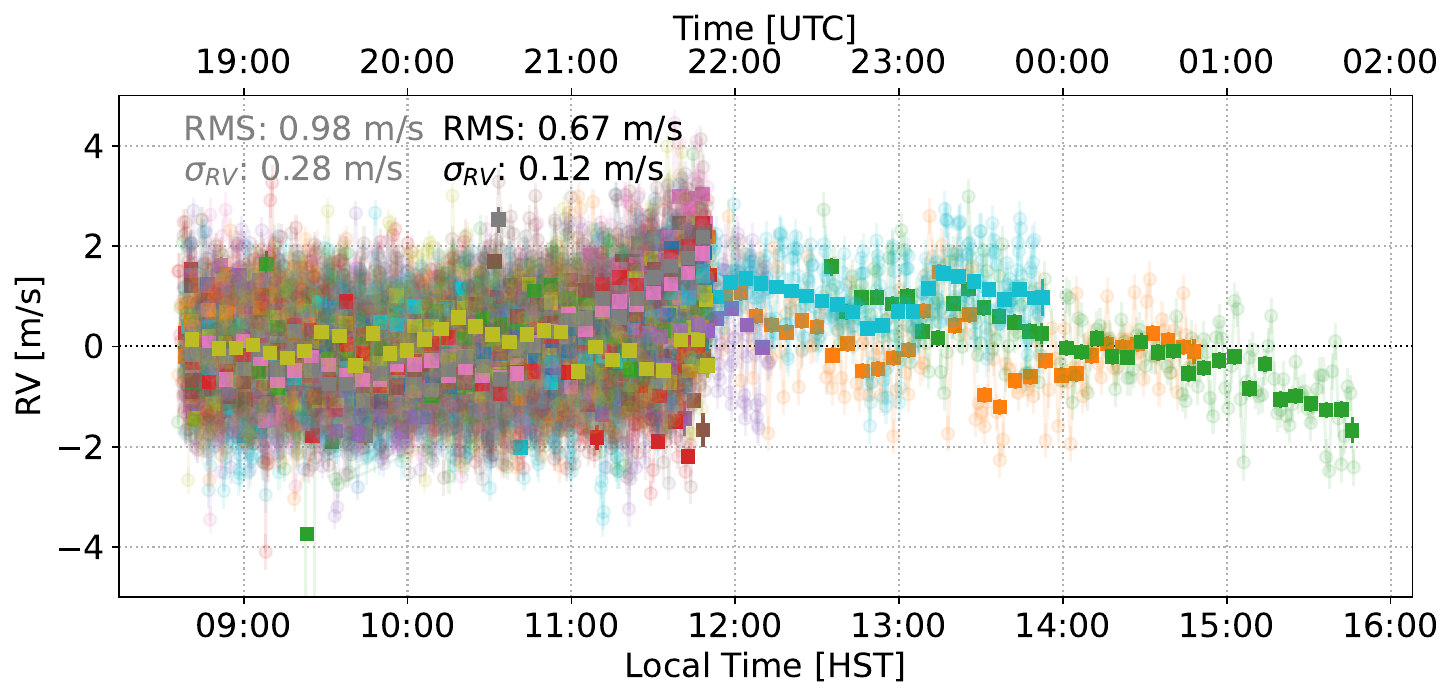}
    \includegraphics[height=0.25\textheight]{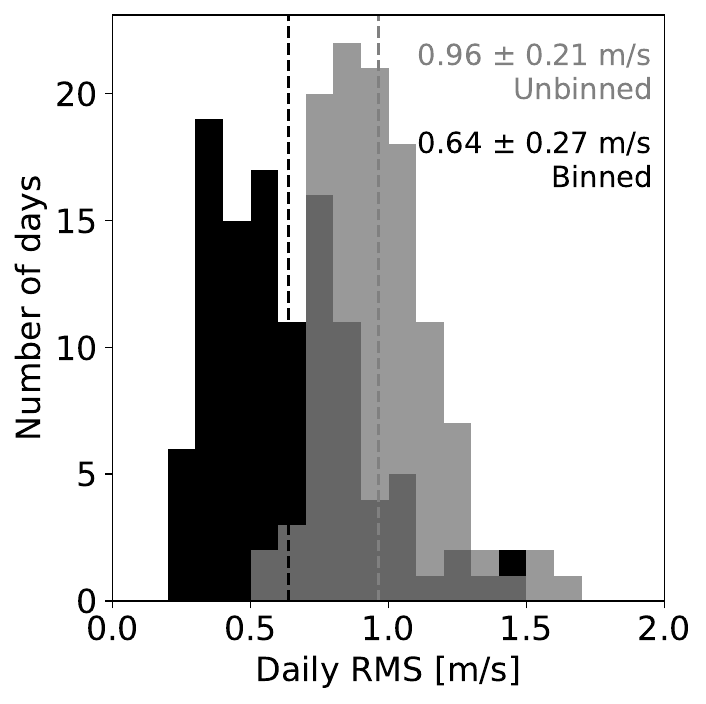}
    \caption{All clear-sky SoCal RVs to date, phased to the time-of-day local time. A daily median value has been subtracted. The raw measured RVs (no drift correction) are shown as faded points, color-coded by day. The bolded points show the same data binned over 5.5~min. The histogram on the right shows the distribution of daily RMS for both the binned and unbinned RVs.}
    \label{fig:daily_bins}  
\end{figure*}

The daily root-mean-squared (RMS) of the RVs after binning over the 5.5~min solar oscillations and correcting for instrumental drift using the simultaneous calibration (etalon) is typically around $0.64 \pm 0.27$~m~s$^{-1}$ (Figure~\ref{fig:daily_bins}). Instrument drift over a daily SoCal sequence (3--6~hrs) is typically below the 1~m~s$^{-1}$ level, although some days show stronger deviations. 

\subsection{Charge Transfer Inefficiency Issue}\label{sec:cti}

\begin{figure*}
    \centering
    \includegraphics[width=0.95\textwidth]{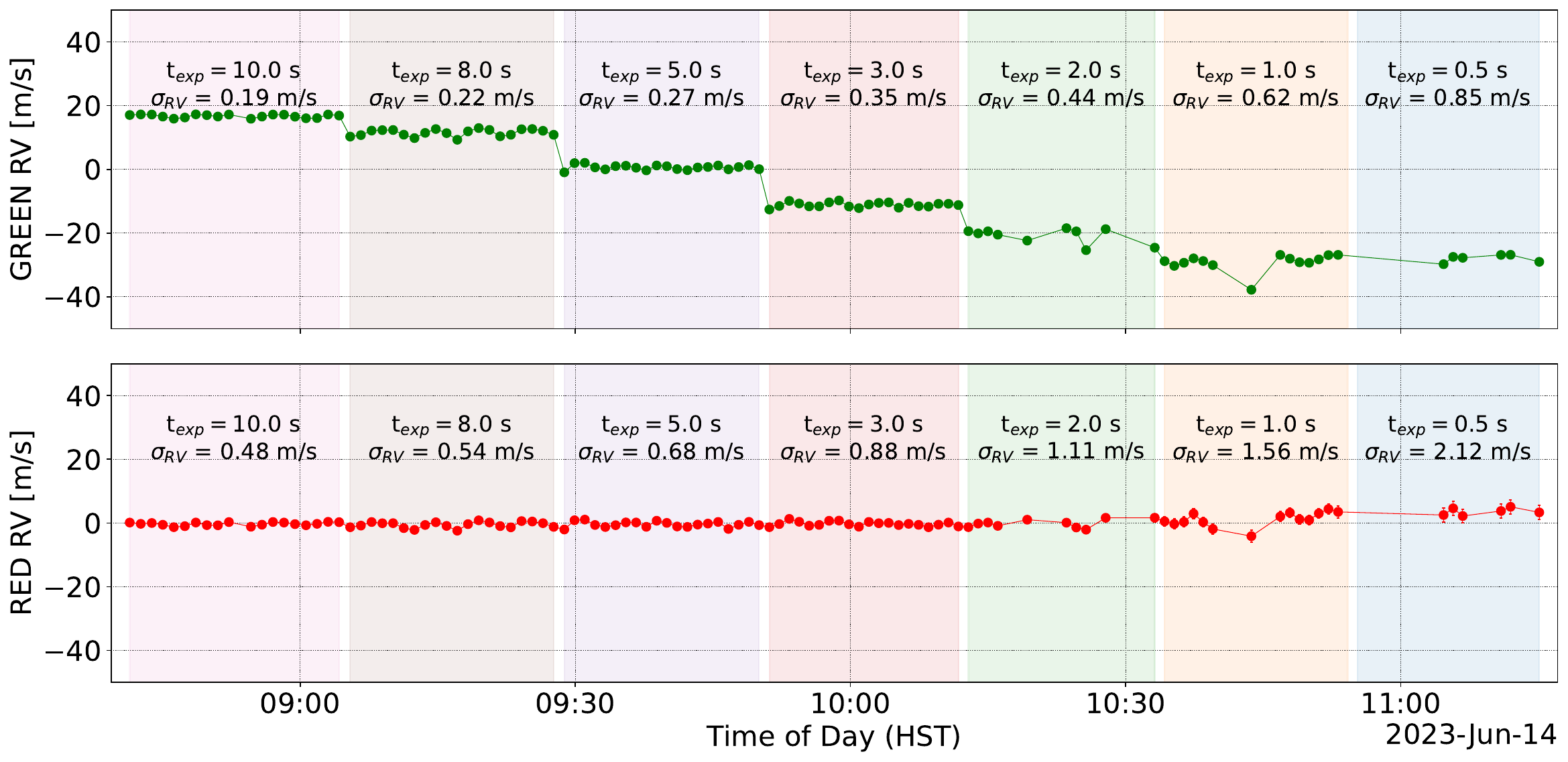}
    \includegraphics[width=0.95\textwidth]{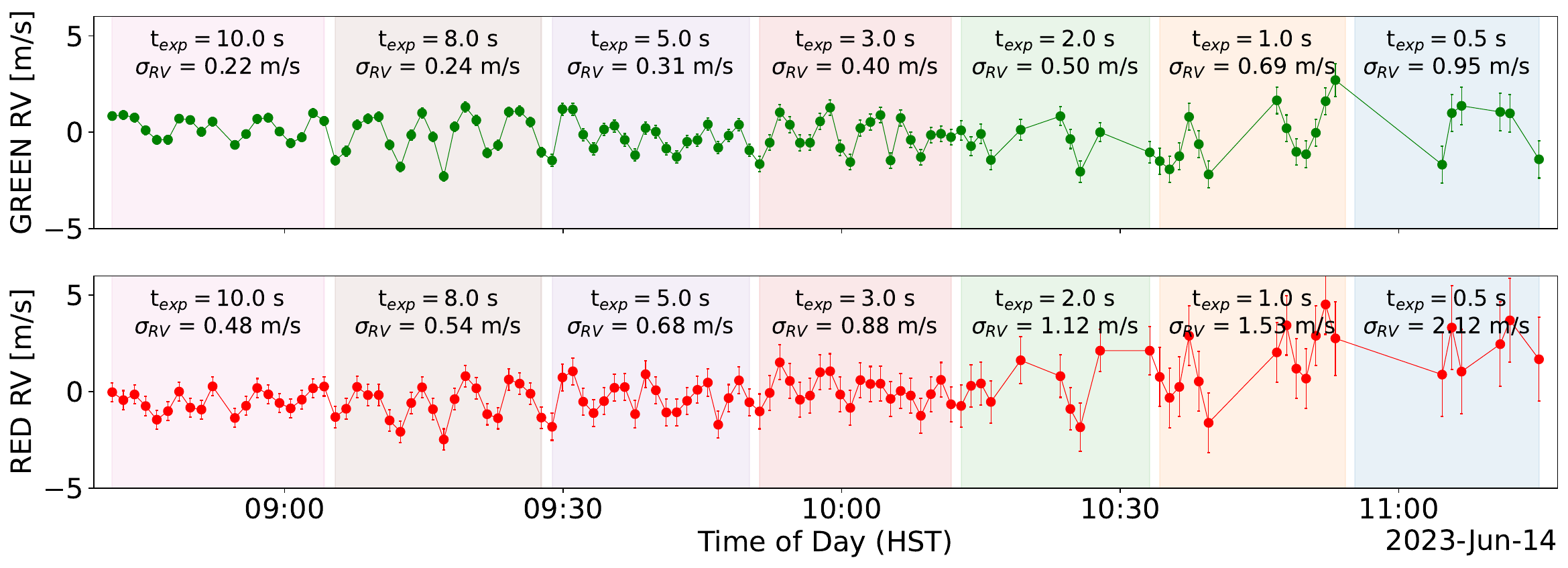}
    \includegraphics[width=0.95\textwidth]{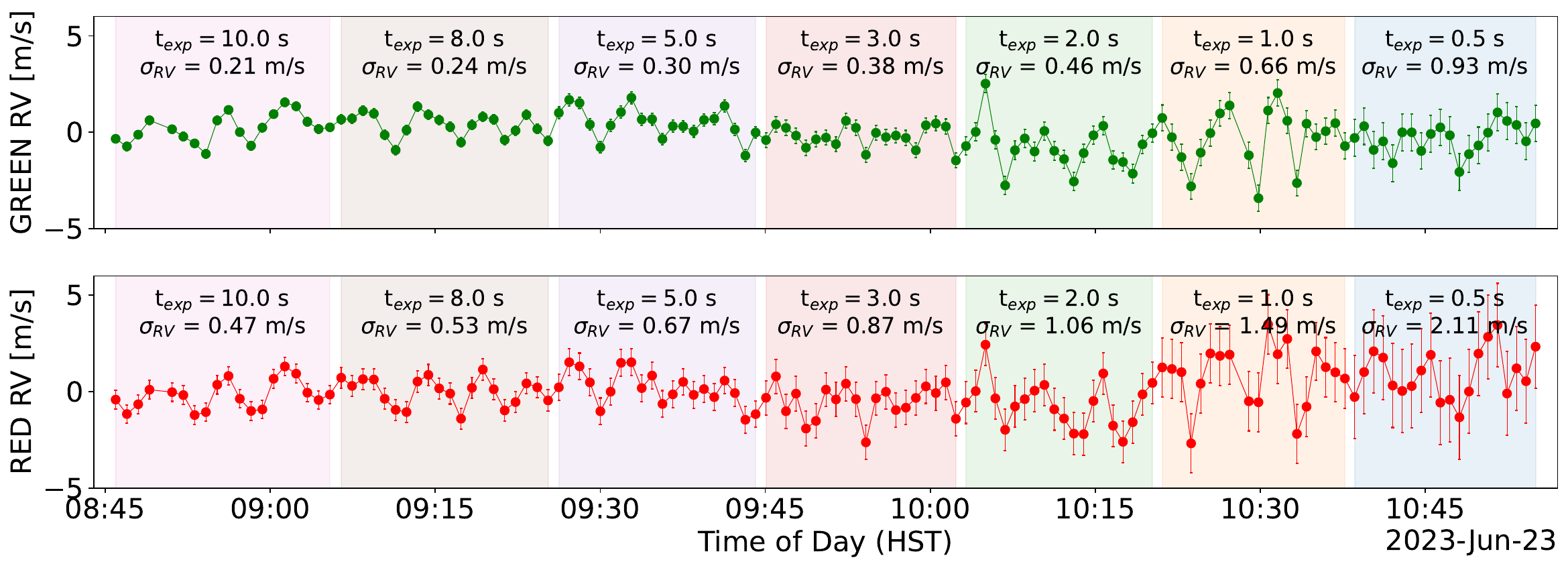}
    \caption{\textbf{Top:} SoCal RVs (green and red) during our CTI test stepping across a range of exposure times. The large offsets between each sequence in the green RVs are caused by CTI effects in one of the four amplifiers. Some gaps exist due to intermittent clouds. \textbf{Middle:} The same data but recomputed by masking the quadrant of the green CCD that is read by the affected amplifier. The offsets disappear below the instrumental noise, at the expense of slightly worse RV precision since over 1/4 of the spectrum in the green channel is not used.
    \textbf{Bottom:} The same sequence of exposure times taken on a different day using a 2-amplifier readout scheme. By not using the affected amplifier, the CTI effects disappear and full RV precision is maintained.
    }
    \label{fig:cti}
\end{figure*}

Another KPF commissioning activity was to measure the impact of charge transfer inefficiency (CTI) in the CCDs on stellar RVs. CTI can produce a SNR-dependent RV shifts since spectral lines will become skewed by the leftover charge smearing across the detector~\citep{Bouchy2009CTI, Halverson2016, Blake2017}. To directly probe the effects of CTI on the KPF RVs, we gathered sequences of 20 exposures at exposure times of 10~sec, 8~sec, 5~sec, 3~sec, 2~sec, 1~sec, and 0.5~sec (see Figure~\ref{fig:cti}). We noticed significant systematic jumps in RV between each sequence. We isolated this effect to one of the four amplifiers on the green CCD by observing that this effect was only present in RVs computed using that quadrant of the 2D spectrum. We measured that this amplifier has roughly 100 times worse CTI than what was measured during laboratory CCD tests performed at Caltech (prior to shipping KPF to Hawai\okina i in the summer of 2022). 

To work around the CTI problem affecting one amplifier on the Green CCD, we developed a new read mode of KPF that utilizes two low-CTI amplifiers operating at 200 kHz in place of the original 4-amplifier, 100 kHz mode. This is the new ``standard'' readout mode of KPF as of June 24, 2023. Note that the fast readout mode still requires all four amplifiers operating at 400 kHz. As a result, all fast readout data as well as all standard readout data prior to June 24, 2023 must have the affected quadrant of the green CCD  masked when computing RVs. This masking is now automatically applied to all previously-collected data in the standard KPF DRP and does successfully resolve the CTI issue (see bottom panel of Figure~\ref{fig:cti}), at the cost of slightly degraded RV precision since over a quarter\footnote{The CTI-affected quadrant is the bluest end of the green detector, where the inter-order spacing is smallest.} of the spectrum is being ignored. We are considering raising the exposure time to 10--12~s when using fast readout mode to compensate for this. This would result in a 26~s cadence, slightly better duty cycle (38\%), and $\sim$800 spectra per 6~hr day, but would reach 25~\cms{} photon-limited precision vs. 36~\cms{} in a 5~sec exposure. For comparison, in standard read mode we can reach $\sim 20$~\cms{} in 10~sec or $28$~\cms{} in 5~sec. Longer exposure times would also fully utilize KPF's unique ability to obtain high SNR spectra; SNR $\sim$1400 \textit{per trace} is reached in the red channel for a 12~sec exposure, and nonlinear CCD response only begins to set in above SNR $\sim$1500 per trace.

\subsection{Comparison to Other EPRV Solar Feeds}\label{sec:neid-comparison}

\begin{figure*}
    \centering
    \includegraphics[width=0.4950\textwidth]{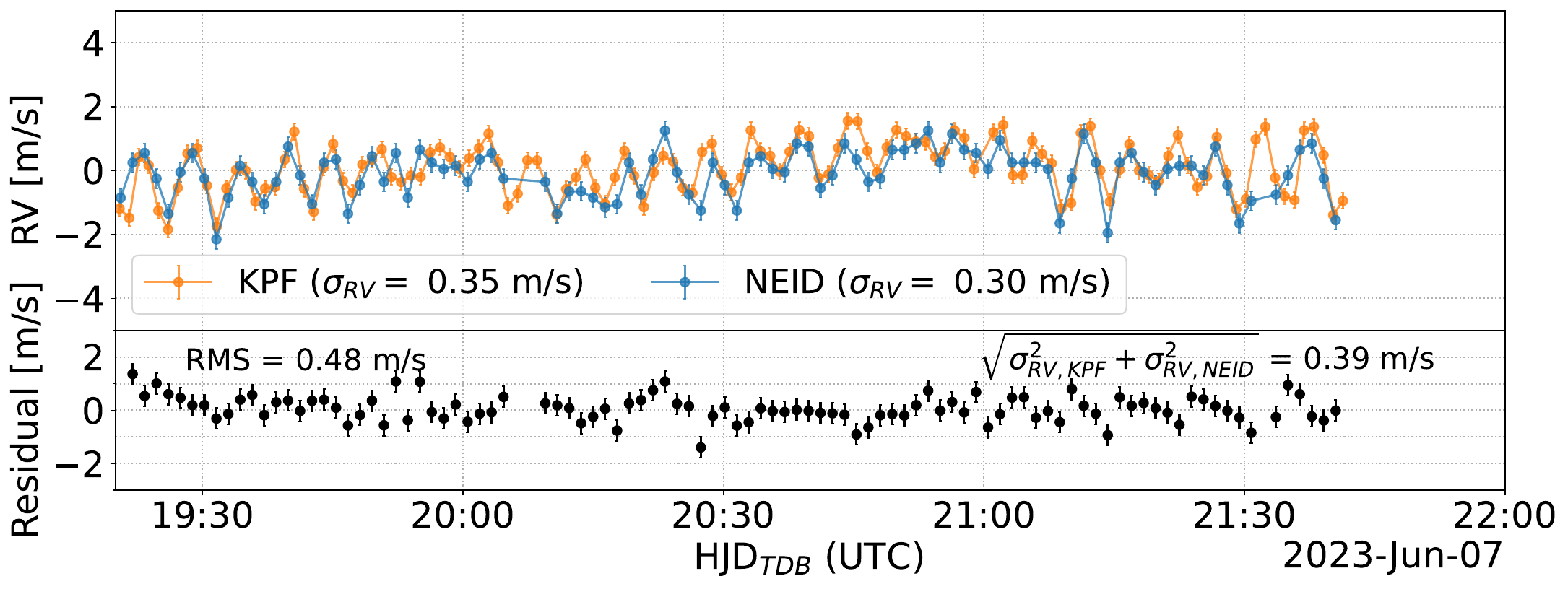}
    \includegraphics[width=0.4950\textwidth]{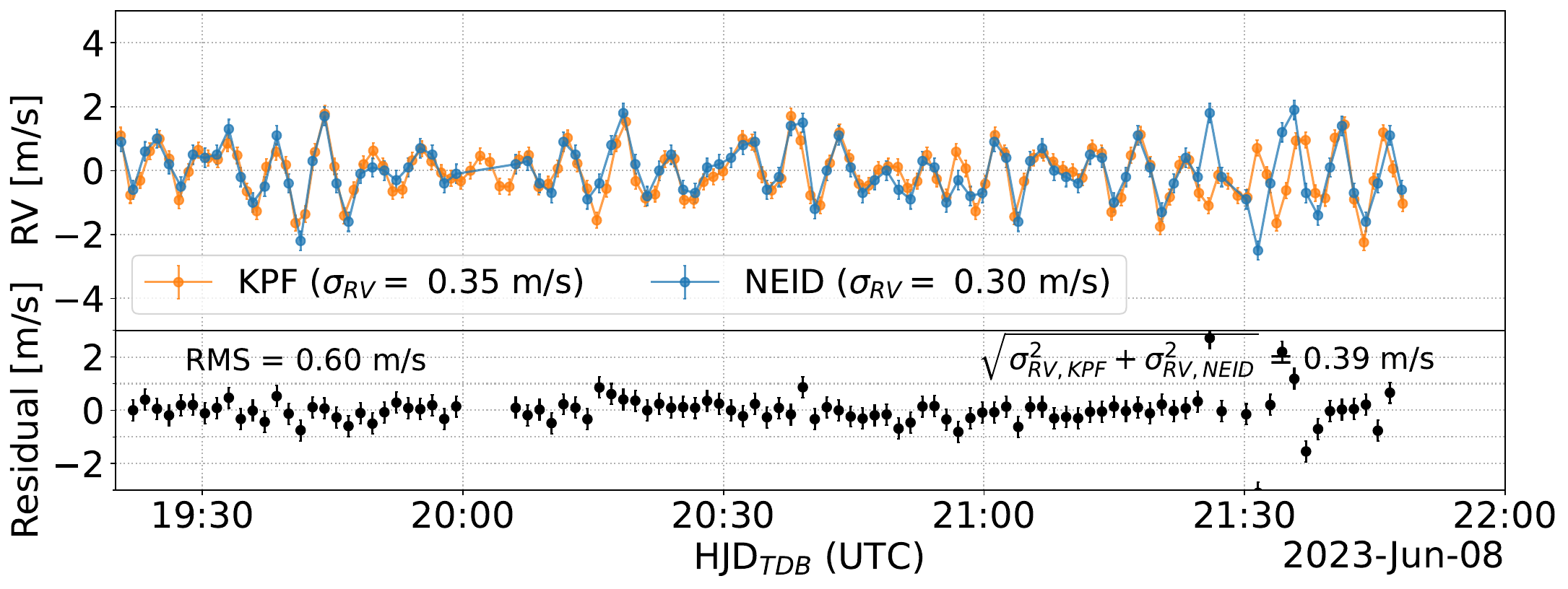}
    \includegraphics[width=0.4950\textwidth]{figures/daily_neid_comparisons/kpfsocal_neidsolar_comparison_2023-06-08.pdf}
    \includegraphics[width=0.4950\textwidth]{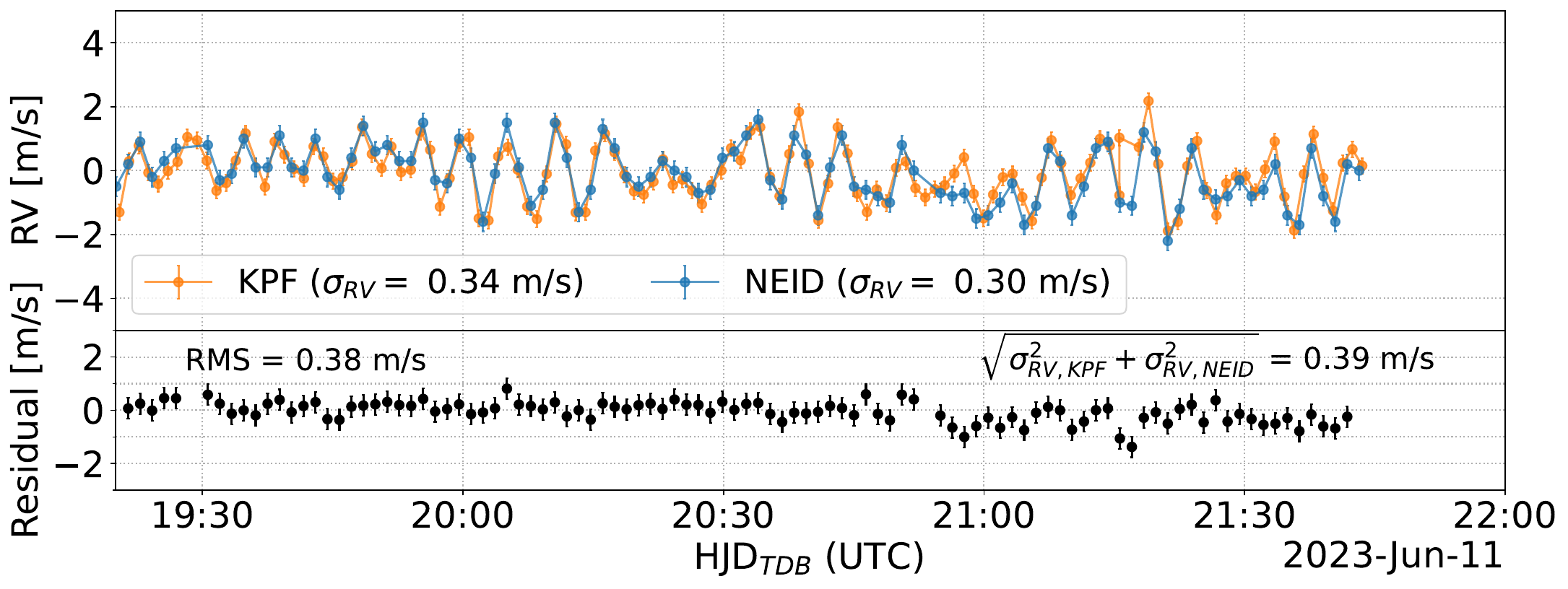}
    \includegraphics[width=0.4950\textwidth]{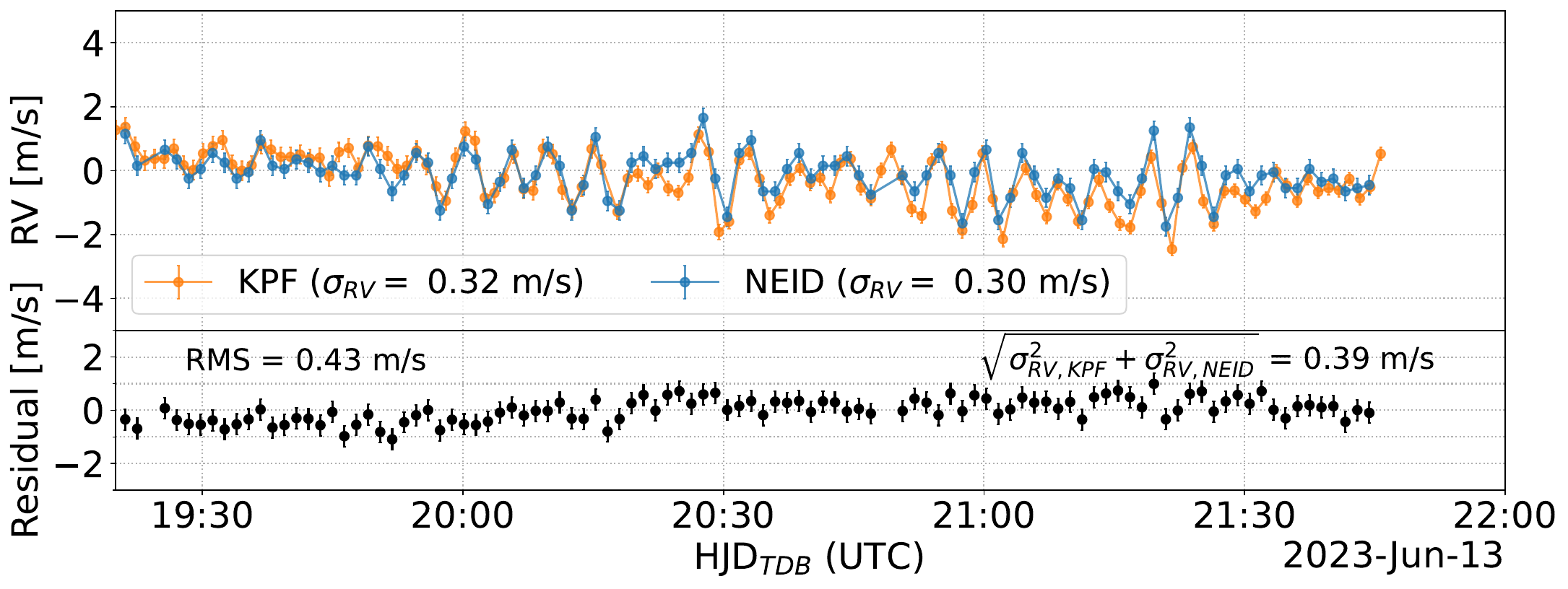}
    \includegraphics[width=0.4950\textwidth]{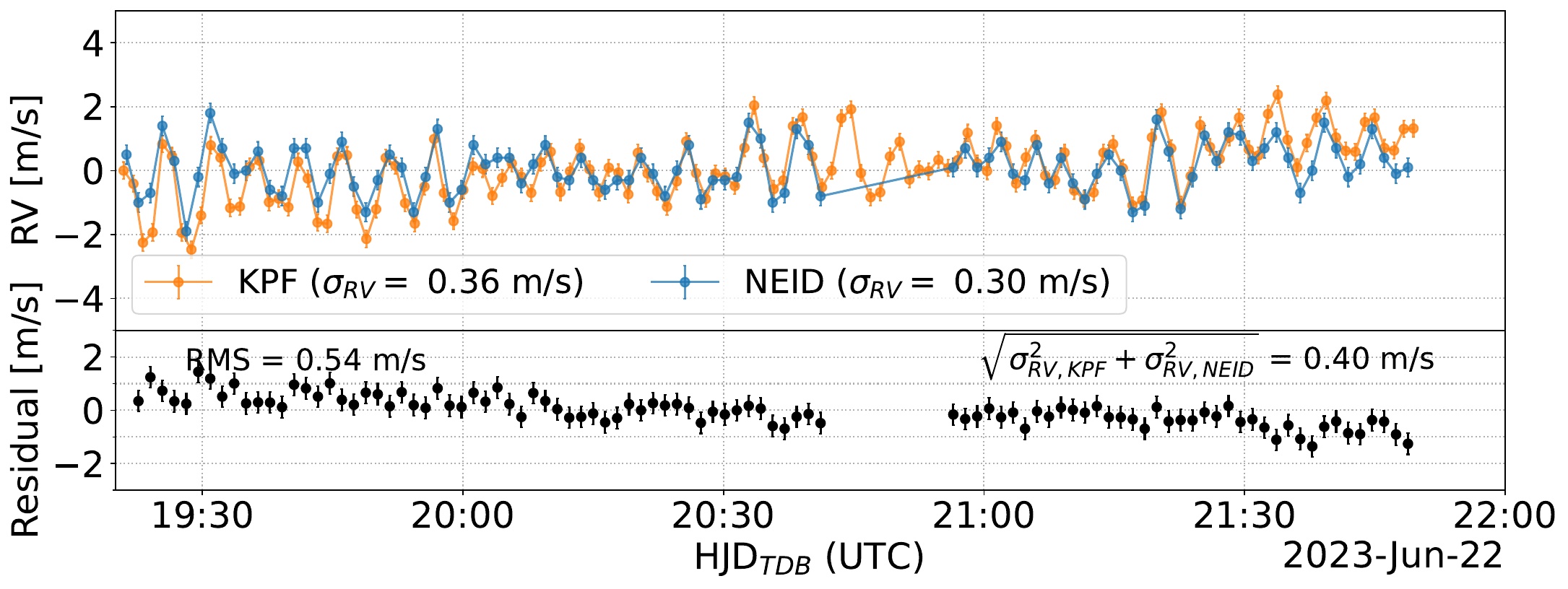}
    \includegraphics[width=0.4950\textwidth]{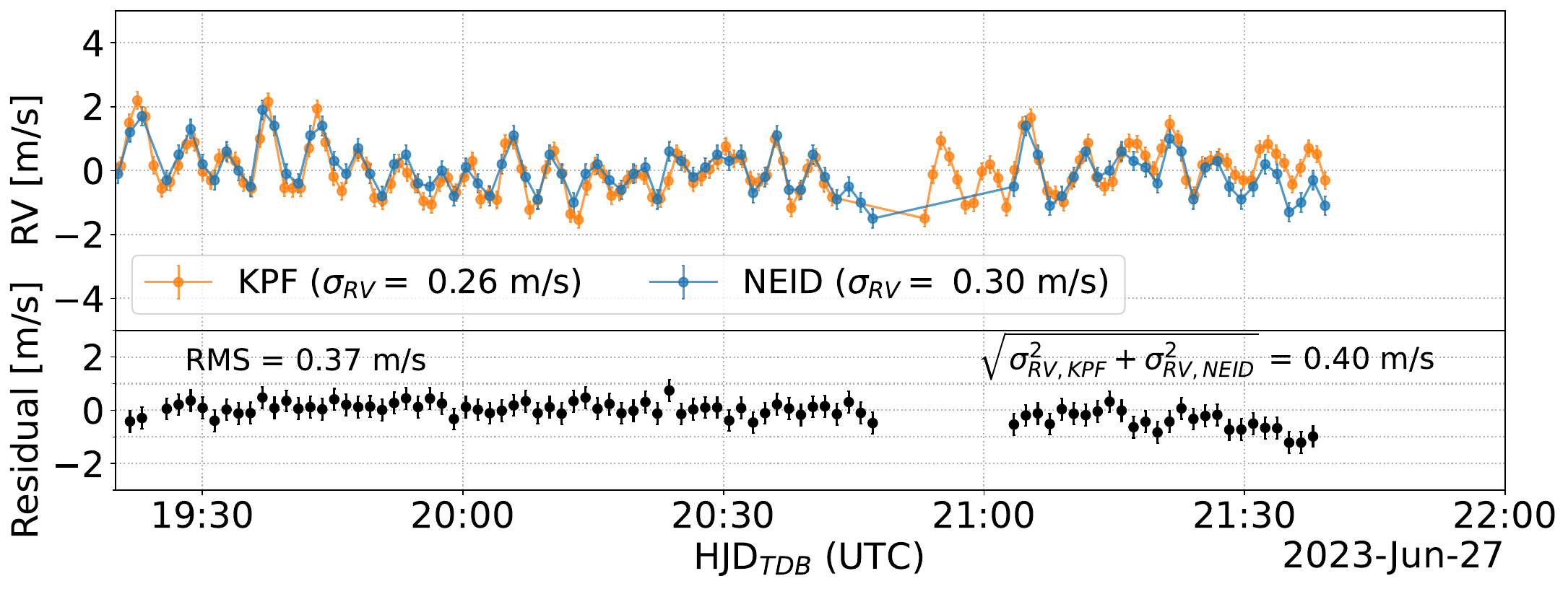}
    \includegraphics[width=0.4950\textwidth]{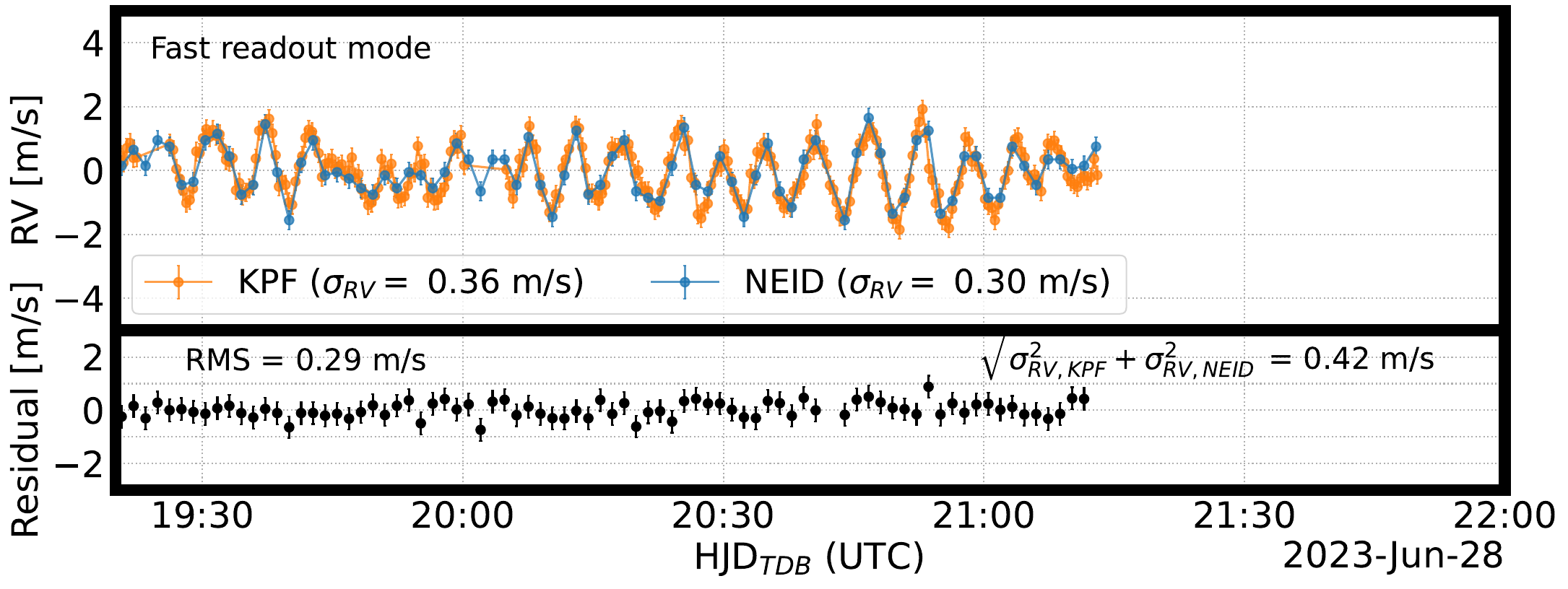}
    \includegraphics[width=0.4950\textwidth]{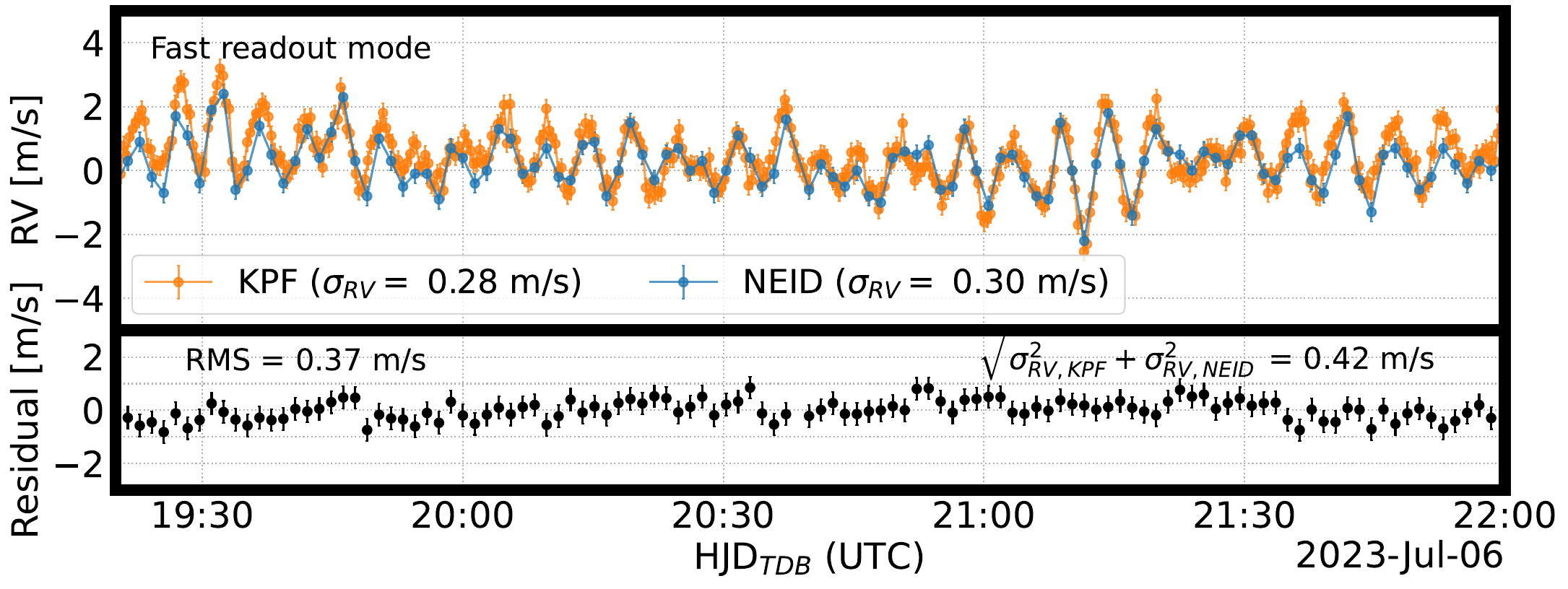}
    \includegraphics[width=0.4950\textwidth]{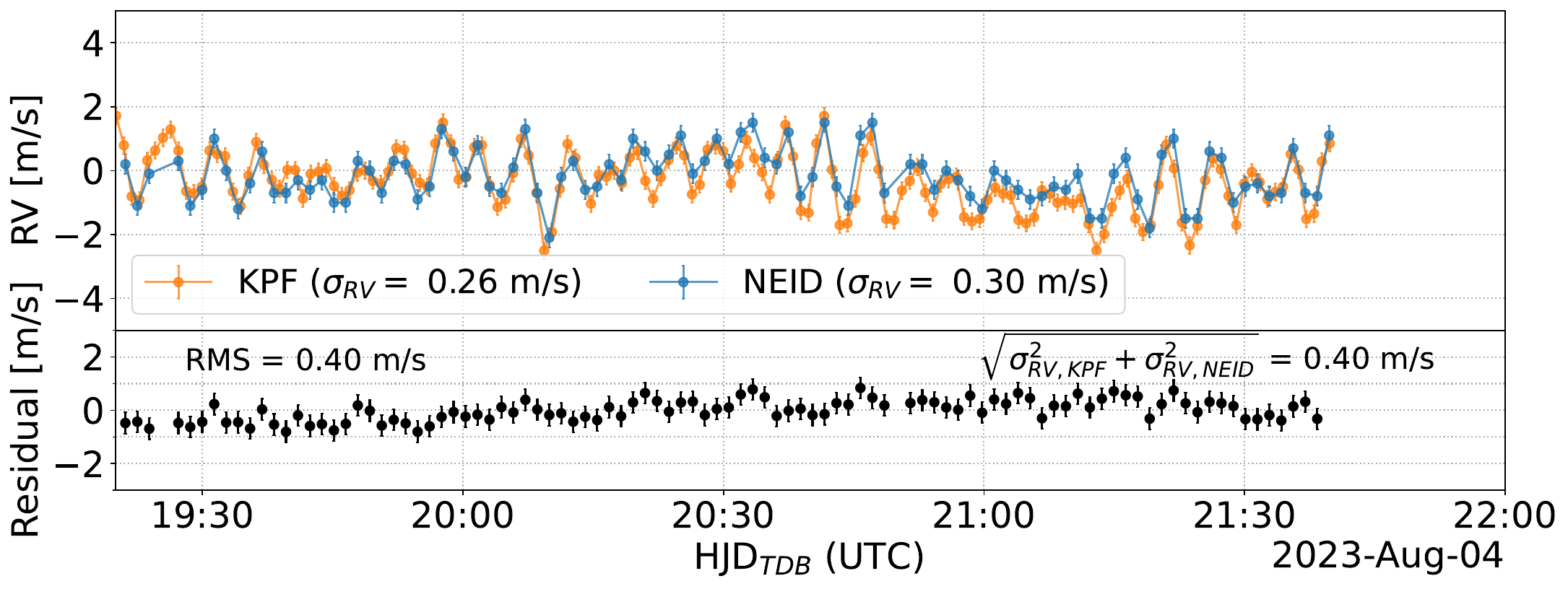}
    \caption{Solar RVs measured by KPF (corrected for drift) and NEID for a selection of days where both sites had clear weather conditions and a drift correction was possible for KPF using the simultaneous calibration. KPF data (orange points) on two of the days, June 28 and July 6, were taken in the fast readout mode (bolded frames), with the rest of the days taken in standard readout mode. The NEID RVs are shown in blue. The 5.5~minute solar p-mode oscillations are clearly observed by both instruments at the same amplitude and phase. The lower panel of each plot shows the residuals between a spline-interpolation of the KPF RVs, sampled at the NEID timestamps, and the NEID RVs. The residual RMS is comparable to the combined instrumental noise floor for most days; some days show a smaller RMS than the combined noise floor. On some days, such  as June 22, the RVs disagree near UT 21:30. This is likely caused by additional instrumental drift in the KPF RVs due to liquid nitrogen fills around HST 11:00 (UT 21:00) not being fully removed by the simple drift model.}
    \label{fig:kpf_neid_comparison}
\end{figure*}

While most active EPRV solar feeds have opted to publish their data in large data releases \citep[e.g.,][]{CollierCameron2019, Dumusque2021}, the NEID Solar Feed makes its data available to the public immediately after it is acquired and reduced\footnote{Available at \href{https://neid.ipac.caltech.edu/search_solar.php}{https://neid.ipac.caltech.edu/search\_solar.php}}. We prioritized morning observations with SoCal (08:45 -- 12:00 HST) as this window fully overlaps with the early afternoon NEID solar observations in Arizona. This way we could immediately compare RVs between instruments.

Figure~\ref{fig:kpf_neid_comparison} shows the measured SoCal (orange) and NEID (blue) RVs for ten days with fully clear skies at both sites. We observed the majority of these days with KPF in the standard readout mode (5~sec exposure, 55~sec cadence), with tests of the fast readout mode  (5~sec exposure, 21~sec cadence) on June 28, 2023 and July 6, 2023. Since the fast readout data are taken in 4-amplifier mode, the RVs are computed using an order mask on the green CCD to avoid contamination by CTI effects (see Section~\ref{sec:cti}), hence the larger than usual per-measurement uncertainty. The NEID solar RVs have a longer exposure time (55~sec) but an intermediate readout time (28~sec), resulting in a similar cadence (83~sec cadence) as our standard read mode data. Both instruments clearly resolve the 5.5~minute solar p-mode oscillations, which dominate the common RV variability on these $\sim$hours intra-day timescales \citep[see][]{Kjeldsen2008}. The bottom panel of each daily plot shows the residuals between the NEID RVs and a spline fit of the KPF RVs interpolated to the NEID timestamps. The RMS of these residuals is typically around 30--40~{\cms}, which is slightly lower than the quadrature-sum of the KPF and NEID single-measurement errorbars (40--46~{\cms}). As the KPF RVs are corrected for instrumental drift using the simultaneous etalon RVs, this means that there are no other sources of unaccounted instrumental noise in these data. For any observations that show large disagreements (e.g. June 8 near 21:30 UTC), a deeper investigation is warranted to isolate which instrument the source of disagreement is coming from. As the KPF DRP, wavelength solutions, calibration source RVs, and drift models are still converging on a long-term stable solution, we leave this investigation for future work when the KPF RVs reach the same level of maturity as the NEID RVs.

The fact that the ``out-of-the-box'' KPF RVs line up so well with the NEID RVs over daily timescales is extremely encouraging. Drift on these timescales for KPF is $< 0.5~\text{\ms}$ $\text{hr}^{-1}$, so we expect similar levels of agreement on longer timescales once the day-to-day offsets between KPF wavelength solutions become sub-{\ms}. Future work expanding on the investigation conducted by \citet{Zhao2023}, who studied one month of overlap between solar RVs from HARPS, HARPS-N, EXPRES, and NEID, will be especially fruitful. Additionally, SoCal will observe the Sun for an additional 2--3 hours after the Sun has set in Arizona for EXPRES and NEID, meaning these five instruments will collect nearly 20 hours of continuous solar RVs in the summer months and $\sim$17 hours in the winter months. By cross-calibrating instruments using the overlapping windows of solar observations, longer-term variability like granulation will be better resolved. However, each instrument adopts a unique observing strategy. HARPS-N takes $\sim$5~min exposures to average over p-modes, EXPRES uses an adaptive exposure time to reach a fixed SNR threshold (typical exposures are around $3$~min), and HARPS and NEID both use short fixed exposure times of $30$~sec and $55$~sec respectively. To compare RVs on longer timescales, these RVs must be binned to shared ``exposure times'' and timestamps, which introduces some uncertainty. The faster cadence of KPF (5~sec exposure and 15~sec readout) directly traces the p-mode oscillations, thus the KPF RVs can be binned to these shared exposure times and timestamps with less inherent error \citep{Zhao2023}. Long term, the publicly available SoCal and NEID RVs will provide crucial benchmarks for understanding instrument performance and for isolating solar activity signals.

\section{Conclusions and Future Work}
\label{sec:conclusion}

We have developed, built, and installed the Solar Calibrator for KPF at W. M. Keck Observatory. SoCal makes use of proven, off-the-shelf components and is protected from extreme weather by a rugged motorized enclosure. Daily operations are performed autonomously with little-to-no human intervention required. We achieved first light on April 25, 2023 and have been observing the Sun almost daily since June 2023, accumulating over 19,000 solar spectra at the time of submitting this manuscript (October 18, 2023). SoCal obtains SNR $\sim 1200$ solar spectra in a 5~sec exposure. When paired with KPF's fast readout mode we are able to record solar RV time series at 21~sec cadence with $< 30$~{\cms} photon-limited precision. Long-term operations can further utilize KPF's high SNR capabilities to acquire spectra with SNR as high as $\sim$2400.

On short timescales, SoCal is demonstrating the EPRV capabilities of KPF extremely well. With no drift correction, binning over the p-mode oscillations reduces the RMS of observed solar RVs to just 20--30~\cms{} on days with minimal instrumental drift and 67~\cms{} across all days. We compared solar RVs from SoCal to those taken simultaneously with NEID and found excellent agreement within individual days; the residual RV between KPF and NEID was comparable to their combined photon-limited precision ($\sim$40~{\cms}). 

Long-term performance validation still requires improvements to the KPF DRP, particularly the stability of daily wavelength solutions, but preliminary results are encouraging. SoCal has also enabled independent monitoring of instrumental drift and will become even more so once comparisons with NEID on longer timescales become possible. This has been especially valuable during times when the LFC was not working and the etalon lamp was degrading. SoCal data was also instrumental in discovering and diagnosing the CTI issue in the KPF detectors as well as exercising and improving the DRP throughout commissioning.

Continued monitoring of the Sun by EPRV facilities across the globe will not only allow for multi-instrument comparisons and calibrations (such as in \citealt{Zhao2023}), but will also provide near-continuous solar monitoring which may help constrain granulation effects. Additionally, the Sun is currently increasing in activity towards solar maximum (\citealt{Upton2023} estimate the peak in fall 2024), making forthcoming cross-instrument studies especially opportune for probing the effects of active features such as spots/faculae/plages on EPRV data. The fast cadence and high SNR of SoCal data allow for more precise binning over short-term oscillations enabling more effective comparisons to other instruments. Soon, the solar feed for MAROON-X will come online. As Gemini-N and WMKO share the same observing conditions (and the same Sun), comparisons between SoCal and MAROON-X solar data will be uniquely advantageous as the only variable is the instrument. Lastly, SoCal's geographic location fills a large gap in reaching continuous 24-hour coverage using the global network of solar feeds.

It will also be interesting to compare EPRV solar data with solar RVs obtained by dedicated asteroseismology observatories. There are two ground-based global networks of solar observatories performing 24/7 helioseismology, the Global Oscillation Network Group \citep[GONG;][]{GONG} and the
Birmingham Solar Oscillations Network \citep[BiSON;][]{Davies2014, Hale2016}. These facilities use a single spectral line to measure solar RVs and have set the standard for measuring the oscillation frequencies of the Sun \citep{Broomhall2009}. The Stellar Oscillations Network Group \citep[SONG;][]{SONG} is a global network of 1~m telescopes with iodine-cell calibrated spectrographs designed to do asteroseismology with RVs on the nearest and brightest stars. A sun tracker was installed at the Hertzsprung SONG telescope at the Teide Observatory in 2017, which collected {\ms} quality RVs of the Sun at a blazing 4~sec cadence (0.5~sec exposure, 3.5~sec readout) for three months in 2018 \citep{SONGsun}. Our interpretations of our solar EPRV datasets would benefit greatly from collaborations with the heliophysics community and detailed comparisons between our rich datasets.

SoCal data is publicly available on the Keck Observatory Archive. Future studies to develop new spectral activity indicators or activity-invariant RV extraction algorithms will be most fruitful on the high SNR, high cadence, and long-baseline solar time series that SoCal and other similar facilities are producing. Solar EPRV datasets are becoming ever more important not just for understanding, calibrating, and optimizing individual spectrograph performance, but also for paving the way to the data analysis tools needed to uncover exo-Earths in stellar EPRV time series.

\section{Acknowledgements}
    
We gratefully acknowledge the efforts and dedication of the Keck Observatory staff, particularly Maylyn Carvalho, Rick Johnston, Matt Barnett, Derek Park, Jerry Pascua, Steve Baca, Bobby Harrington III, Danny Baldwin, Randy Ching, Hamza Elwir, Ed Wetherell, Justin Ballard, Todd Von Boeckmann, Chris Martins, Daniel Orr, Max Brodheim, and Kyle Lanclos. We thank G\'abor Kov\'acs for designing the enclosure electronics and providing troubleshooting guidance, and Gaspar Bakos for helpful design discussions and for facilitating the acquisition of the enclosure. We thank Kodi Rider for helping to coordinate SoCal operations at SSL in Berkeley. We thank Bradford Holden and William Deich for designing the KPF data structures and helping with FITS/KTL keywords. We thank Andy Monson and Andrea Lin for useful discussions about the tracker assembly, and for providing detailed solid models of key components of the NEID Solar tracker assembly.

We extend our deepest gratitude to the Kahu K\=u Mauna (Guardians of the Mountain), the Center for Maunakea Stewardship’s Environmental Committee, and the Maunakea Management Board for their thoughtful review and approval of the SoCal project permit. The summit of Maunakea is a place of significant ecological, cultural, and spiritual importance within the indigenous Hawaiian community. We understand and embrace our accountability to Maunakea and the indigenous Hawaiian community, and commit to our role in long-term mutual stewardship.

R.A.R. acknowledges support from the National Science Foundation through the Graduate Research Fellowship Program (DGE 1745301). The Solar Calibrator was supported in part by the Heising-Simons Foundation through grant 2022-3931, the Simons Foundation grant ``Planetary Context of Habitability and Exobiology,'' and the Suzanne \& Walter Scott Foundation.

Some of the data presented herein were obtained at the W. M. Keck Observatory, which is operated as a scientific partnership among the California Institute of Technology, the University of California, and the National Aeronautics and Space Administration. The Observatory was made possible by the generous financial support of the W. M. Keck Foundation. This research has made use of the Keck Observatory Archive (KOA), which is operated by the W. M. Keck Observatory and the NASA Exoplanet Science Institute (NExScI), under contract with the National Aeronautics and Space Administration. The research was carried out, in part, at the Jet Propulsion Laboratory, California Institute of Technology, under a contract with the National Aeronautics and Space Administration (80NM0018D0004).

\facility{Keck Planet Finder, W. M. Keck Observatory}

\software{
\texttt{astropy}    \citep{astropy},
\texttt{astroplan}  \citep{astroplan},
\texttt{barycorrpy} \citep{barycorrpy},
\texttt{matplotlib} \citep{matplotlib},
\texttt{pvlib}      \citep{pvlib},
\texttt{pymodbus}, 
\texttt{pytransitions}, \citep{pytransitions},
\texttt{scipy}      \citep{scipy},
\texttt{socket}, 
\texttt{websockets} 
}

\bibliography{main}
\bibliographystyle{aasjournal}


\end{document}